\documentstyle[aaspp4]{article}

\lefthead{MENOU et al.}
\righthead{BH AND NS TRANSIENTs IN QUIESCENCE}

\def\spose#1{\hbox to 0pt{#1\hss}}
\def\lta{\mathrel{\spose{\lower 3pt\hbox{$\mathchar"218$}}
     \raise 2.0pt\hbox{$\mathchar"13C$}}}
\def\gta{\mathrel{\spose{\lower 3pt\hbox{$\mathchar"218$}}
     \raise 2.0pt\hbox{$\mathchar"13E$}}}

\begin{document}

\title{Black Hole and Neutron Star Transients in Quiescence}

\author{Kristen Menou,\altaffilmark{1,2} Ann A. Esin,\altaffilmark{3,4}
Ramesh Narayan,\altaffilmark{1}\\ 
Michael R. Garcia,\altaffilmark{1} Jean-Pierre Lasota\altaffilmark{2,5} \& Jeffrey E. McClintock\altaffilmark{1}}

\altaffiltext{1}{Harvard-Smithsonian Center for Astrophysics, 60
Garden Street, Cambridge, MA 02138, USA.\\  kmenou@, rnarayan@,
mgarcia@, jmcclintock@cfa.harvard.edu} 
\altaffiltext{2}{UPR 176 du
CNRS, DARC, Observatoire de Paris, Section de Meudon, 92195 Meudon
C\'edex, France.} 
\altaffiltext{3}{Theoretical Astrophysics,
Caltech 130-33, Pasadena CA 91125, USA. aidle@tapir.caltech.edu}
\altaffiltext{4}{AXAF Fellow}
\altaffiltext{5}{Present address: Institut
d'Astrophysique de Paris, 98bis Boulevard Arago, 75014 Paris, France.\\ lasota@iap.fr}

\begin{abstract}
  We consider the X-ray luminosity difference between neutron star and
  black hole soft X-ray transients (NS and BH SXTs) in quiescence.
  The current observational data suggest that BH SXTs are
  significantly fainter than NS SXTs.  The luminosities of quiescent
  BH SXTs are consistent with the predictions of binary evolution
  models for the mass transfer rate if (1) accretion occurs via an
  ADAF in these systems and (2) the accreting compact objects have
  event horizons.  The luminosities of quiescent NS SXTs are not
  consistent with the predictions of ADAF models when combined with
  binary evolution models, unless most of the mass accreted in the
  ADAF is prevented from reaching the neutron star surface. We
  consider the possibility that mass accretion is reduced in quiescent
  NS SXTs because of an efficient propeller and develop a model of the
  propeller effect that accounts for the observed luminosities.  We
  argue that modest winds from ADAFs are consistent with the
  observations, while strong winds are probably not.
\end{abstract}


\section{Introduction}

Soft X-ray Transients (SXTs) are compact binary systems in which a
low-mass secondary (either a main-sequence star or a subgiant)
transfers mass via Roche-lobe overflow onto a black hole (BH) or
neutron star (NS) primary (see reviews by Tanaka \& Lewin 1995; van
Paradijs \& McClintock 1995; White, Nagase \& Parmar 1995).  SXTs have
highly variable luminosities.  They spend most of their lifetimes in a
low luminosity quiescent state, but occasionally undergo dramatic
outbursts during which both the optical and X-ray emission increase by
several orders of magnitude (e.g. Chen, Shrader \& Livio 1997;
Kuulkers 1998). NS SXT outbursts typically occur every 1-10 years and
last for several weeks, while BH SXT outbursts are typically
separated by 10-50 years (or perhaps longer) and last for several
months (see Chen et al. 1997).

A variety of observations (see e.g. Tanaka \& Shibazaki 1996) indicate
that, near the peak of an outburst, SXTs accrete matter via a standard
thin disk (Shakura \& Sunyaev 1973), so that there is little doubt
that the accretion is radiatively efficient during this phase. The
situation is more complex in quiescence. The spectra of quiescent BH
SXTs do not resemble that of a thin disk, and the accretion rates
inferred from the observed X-ray luminosities disagree by orders of
magnitude with the predictions of the standard disk instability model
for quiescent disks (e.g. Lasota 1996).

Narayan, McClintock \& Yi (1996) and Narayan, Barret \& McClintock
(1997) showed that the observations of quiescent BH SXTs can be
explained by a two-component accretion flow model consisting of an
inner hot advection-dominated accretion flow (ADAF; Narayan \& Yi
1994, 1995a; Abramowicz et al. 1995; Ichimaru 1977; see Narayan,
Mahadevan \& Quataert 1998 and Kato, Mineshige \& Fukue 1998 for
reviews of ADAFs) surrounded by an outer thin disk.  In the recent
version of this model described in Narayan et al. (1997a), only the
inner ADAF contributes to the observed optical, UV and X-ray emission
of the system.  The outer thin disk acts mainly as a reservoir which
accumulates mass until the next outburst is triggered. In quiescence,
the emission of the disk is primarily in the infrared; this radiation
is hardly seen as it is swamped by the emission of the secondary (see
Narayan et al. 1996, 1997a and Lasota, Narayan \& Yi 1996 for
details).

A key feature of the Narayan et al. (1996, 1997a) model of quiescent
BH SXTs is the low radiative efficiency of the ADAF.  In these flows,
the bulk of the viscously dissipated energy is stored in the gas and
advected with the flow into the black hole (Narayan \& Yi 1995b;
Abramowicz et al. 1995; Narayan et al. 1996, 1997a).  This explains
the unusually low luminosities of BH SXTs. By contrast, in NS SXTs all
the advected energy is expected to be radiated from the neutron star
surface, resulting in a much higher radiative efficiency of the
accretion flow even in the presence of an ADAF (Narayan \& Yi 1995b).
Motivated by this fundamental distinction between black hole and
neutron star systems, Narayan, Garcia \& McClintock (1997) and Garcia
et al.  (1998) compared the outburst amplitudes of BH SXTs and NS SXTs
as a function of their maximum luminosities and showed that the
observations reveal systematically lower relative luminosities in BH
SXTs. They argued that this constitutes a confirmation of the presence
of an event horizon in BH SXTs.  The argument was challenged by Chen
et al. (1998).

In this paper, we attempt to develop a physical understanding of the
difference in quiescent luminosities between NS and BH SXTs. We first
show, in \S 2, that there is indeed a significant difference in the
observed quiescent luminosities of the two classes of objects, in
contrast to the claim of Chen et al. (1998).  We then use binary
evolution models in \S 3 to estimate mass transfer rates in SXTs. We
combine these estimates with the ADAF+thin disk accretion scenario,
taking into account the presence of an event horizon in BH systems and
a reradiating surface in NS systems, to determine the expected X-ray
luminosities in quiescence (\S 4).  The model predictions agree well
with the observations of quiescent BH SXTs. However, the model
substantially overestimates the luminosities of quiescent NS SXTs.  In
\S 5 we show that the NS SXT data can be reconciled with the
predictions of the model if we take into account the ``propeller
effect'' (Illarionov \& Sunyaev 1975), whereby the magnetosphere of a
rapidly rotating neutron star prevents much of the accreting material
from reaching the surface of the neutron star.  In \S 6 we show that
an ADAF model with a moderate wind and a somewhat less efficient
propeller (in NS SXTs), is also consistent with the observed quiescent
luminosities of BH and NS SXTs.  Finally, in \S 7 we discuss possible
limitations and extensions of this work and in \S 8 we summarize the
main results.

\section{Observations}

In Table~\ref{tab:systems}, we list key parameters of several NS and
BH SXTs: the orbital period $P_{\rm orb}$, the distance $D$, the
quiescent X-ray luminosity $L_{\rm min}$ in the $0.5-10$ keV X-ray
band, and the mass $m_1$ (in solar units) of the compact primary.
Since our main interest is in the quiescent emission, we list only
systems for which there exist reliable measurements of $L_{\rm min}$.
The values of $L_{\rm min}$ are taken from Narayan et al. (1997b) and
Garcia et al.  (1998), except for the BH SXT H1705-250 (see below).

In selecting the systems listed in Table~\ref{tab:systems}, we were
careful to avoid any possible confusion between BH and NS SXTs, since
this would make a comparison between the two sets of objects less
reliable.  Thus, we limit our sample to the eight BH candidates that
have firm {\it dynamical} lower limits on the mass of the primary
(e.g. McClintock 1998).  (Note that this is not true for several
so-called BH ``candidates'' of Chen et al.  1998, which seriously
weakens their arguments.)  Similarly, there is firm evidence that the
five systems listed in Table~\ref{tab:systems} as NS SXTs contain NS
primaries, based on the detection of type I X-ray bursts (e.g. Narayan
et al. 1997b; Chen et al. 1997).

Here we report a new and improved luminosity limit for H1705-250 (=Nova
Oph 77) based on the ASCA observation of 18 September 1996, which we
extracted from the HEASARC archive.  The exposure time was 31~ks, but
unfortunately the target was near the chip boundaries which limits the
effective area.  The 4$\sigma$ upper limit is $0.9 \times 10^{-3}$
c~s$^{-1}$, which corresponds to an X-ray flux $(1-40~{\rm keV}) < 1
\times 10^{-13}$ergs~s$^{-1}$cm$^{-2}$.  Assuming a distance of
8.6~kpc, $L_x < 0.9 \times 10^{33}$ergs~s$^{-1}$.

Luminosities can be expressed in units of the Eddington luminosity,
\begin{equation}
L_{\rm Edd} = 1.25 \times 10^{38} m_1~{\rm erg~s^{-1}},
\label{eq:ledd}
\end{equation}
where $m_1$, the mass of the compact object in solar units, is
listed in Table~\ref{tab:systems}. For a standard radiative efficiency
of $10$ \%, the Eddington luminosity corresponds to a mass accretion
rate of
\begin{equation}
\dot M_{\rm Edd} = 1.39 \times 10^{18} m_1~{\rm g~s^{-1}}.
\label{eq:mdedd}
\end{equation}

In this section, we use both absolute luminosities and
Eddington-scaled luminosities\footnote{Note that Eddington-scaled
  luminosities are somewhat uncertain if the primary mass $m_1$ in a
  system is not well known (Eq.~[\ref{eq:ledd}]). This is for instance
  the case of the BH SXT J1655-40. Orosz \& Bailyn (1997) estimate
  that $m_1 = 7$ in this system, while Phillips et al. (1999) claim
  that $m_1$ could be as low as $4.1$} since it is not clear which is
the more appropriate quantity for comparisons. Our calculations in \S
3 suggest that the mass transfer rates in NS and BH SXTs are similar
when expressed in units of $\dot M_{\rm Edd}$, especially at short
orbital periods.  Thus, Eddington-scaled luminosities might be more
appropriate to compare the two classes of objects at short periods
(where most of the data lie). However, this argument is not very
strong. Therefore, we also show absolute luminosities, expressed in
units of $10^{38}$ erg s$^{-1}$ (which is roughly the luminosity
expected from an object accreting at $10^{18}$ g s$^{-1}$ with
a radiative efficiency of $10$ \%). Finally, the ratio $L_{\rm min} /
L_{\rm max}$ (which is independent of the distance $D$) is used for
comparison with previous investigations of the luminosity difference
between BH and NS SXTs. The values of $L_{\rm max}$, the outburst peak
luminosity, are taken from Garcia et al. (1998).

In Figure~\ref{fig:lumtot}, we show the quiescent luminosities of the
BH and NS systems listed in Table~\ref{tab:systems} as a function of
their orbital periods $P_{\rm orb}$.  The open circles correspond to
NS SXTs and the filled dots correspond to BH systems.  Luminosity
upper limits (indicated by downward arrows) are shown for five SXTs,
all of which are BH systems. The orbital period of the NS SXT H1608-52
is uncertain (see Table~\ref{tab:systems}) and its location in
Figure~\ref{fig:lumtot} is indicated by dashed circles. Among the
undetected SXTs, we choose to include in our sample only those systems
that have been observed for more than 10 ks (see
Table~\ref{tab:systems} herein, and also Table~1 in Narayan et
al. 1997b), which therefore have flux limits $\lta 3 \times 10^{-13}$ ergs
s$^{-1}$ with current X-ray satellites (0.5-10 keV; ROSAT/ASCA).

The argument for using $P_{\rm orb}$ along the horizontal axis of
Figure~\ref{fig:lumtot} is as follows.  For any binary system with a
low mass secondary, and transferring mass via Roche-lobe overflow, the
density of the secondary essentially determines $P_{\rm orb}$. At a
given $P_{\rm orb}$, a BH SXT and a NS SXT will have similar
secondaries, so that the mass transfer characteristics are likely to
be similar. Thus a reliable comparison of the radiative efficiencies
of their accretion flows would be possible. In contrast, if we were to
compare the quiescent luminosities of a BH SXT and a NS SXT with quite
different $P_{\rm orb}$, a difference in the mass transfer rates (see
\S 3) could mask actual differences in the radiative efficiencies of
the accretion flows. This point, which is the motivation for the
calculations in \S 3, was first emphasized by Lasota \& Hameury
(1998).

Figure~\ref{fig:lumtot} strongly suggests that BH SXTs are fainter
than NS SXTs. This is especially true when the data are plotted in
Eddington units (Fig.~\ref{fig:lumtot}a), but it is reasonably
convincing even in absolute units (Fig.~\ref{fig:lumtot}b) or in terms
of the ratio $L_{\rm min}/L_{\rm max}$ (Fig.~\ref{fig:lumtot}c).  Note
that all five of the NS SXTs have been detected in quiescence and have
well measured values of $L_{\rm min}$, while the majority of BH SXTs
have only upper limits on $L_{\rm min}$. In our view, this is
significant and underlines the effect we are claiming.

To make a clearer comparison between BH and NS SXTs with the existing
data, we isolate in Figure~\ref{fig:lumfirm} those systems for which
$L_{\rm min}$ and $P_{\rm orb}$ are well determined (see also the very
similar Fig.~3 of Lasota \& Hameury 1998). Although there are only a
few systems in this figure, we feel that we can conclude that (at
least in this small sample) BH SXTs in quiescence are less
luminous than NS SXTs. The difference is as much as two orders of
magnitude (in Eddington units) for short period systems.

It is important to stress here the excellent upper limit on the
quiescent luminosity of the BH SXT J0422+32. This upper limit is
nearly equal to the very low measured quiescent luminosity of
A0620-00. Thus, there are now two BH SXTs that are two orders of
magnitude less luminous, in Eddington units, than quiescent NS SXTs
with comparable $P_{\rm orb}$. It is important to carry out more
sensitive observations of those BH SXTs which have only upper limits
currently.

Narayan et al. (1997b) explained the difference in quiescent
luminosities of NS SXTs and BH SXTs as due to the presence of a hard
surface in NS SXTs vs an event horizon in BH SXTs. The presence of a
surface causes the radiative efficiency in NS SXTs to be high, while
in BH SXTs most of the dissipated energy is advected into the event
horizon of the BH, without producing observable emission.  Narayan et
al. argued that this constitutes observational evidence for event
horizons in black holes.  In the remainder of the paper, we expand on
this idea and construct a physical model of accretion in NS and BH
systems to explain quantitatively the observations summarized in
Figure~\ref{fig:lumfirm}.  In \S 3 we estimate the mass transfer rate
in SXTs, and in \S 4 and \S 5 we describe an accretion model that is
able to reproduce the observed emission in the $0.5-10$ keV band.

\section{Mass Transfer Rates Predicted By Binary Evolution Models}

The theory of binary evolution relies on the Roche-lobe model for the
description of mass transfer (Frank, King \& Raine 1992).  Depending
on which mechanism drives the mass transfer, whether it is loss of
orbital angular momentum through gravitational radiation and magnetic
braking, or expansion of the donor as it evolves away from the
main-sequence, the binaries are classified as j-driven or n-driven
systems. There is a `bifurcation' orbital period $P_{\rm bif}$
separating the two classes such that for $P_{\rm orb} > P_{\rm bif}$
we find n-driven systems whose orbital periods increase with time, and
for $P_{\rm orb} < P_{\rm bif}$ we find j-driven systems whose $P_{\rm
  orb}$ decrease with time.  Although there is no unique value of
$P_{\rm bif}$, because it depends on the donor mass and the strength
of angular momentum losses, estimates range from $0.5-2$ days (e.g.
Pylyser \& Savonije 1988; King, Kolb \& Burderi 1996).

The exact form of the magnetic braking law is not well established
(see e.g. Kalogera, Kolb \& King 1998).  Moreover, Menou, Narayan \&
Lasota (1998) showed that there is strong circumstantial evidence that
magnetic braking (MB) is weak in BH SXTs. In the following, for
simplicity, we neglect the influence of MB in BH and NS SXTs.  We will
see in \S 4.2 that any contribution to the mass transfer by MB would
only strengthen the argument that NS SXTs shed mass rather than
accrete it.

Following King et al. (1996), the mass transfer rates in j-driven
systems with gravitational radiation (GR) and n-driven system with
secondary expansion (EXP) is given by:
\begin{eqnarray}
\dot M_T & \approx & 1.27 \times
 10^{14} m_1 m_2^2 (m_1+m_2)^{-1/3} \left ( P_{\rm orb} / 1 {\rm d}
 \right)^{-8/3} {\rm g~s^{-1}},~P_{\rm orb} < P_{\rm bif}~~~{\rm (GR)},\\ 
& \approx & 2.54
 \times 10^{16} m_2^{1.47} \left ( P_{\rm orb} / 1 {\rm d}
 \right)^{0.93} {\rm g~s^{-1}},~P_{\rm orb} > P_{\rm bif}~~~{\rm (EXP)},
\label{ev}
\end{eqnarray}
where $m_2$ is the mass of the secondary in solar units, $P_{\rm orb}$
is scaled in units of 1 day.

The secondary in j-driven binaries is usually a main-sequence star
(see King et al. 1996 for possible complications) filling its Roche
lobe. These stars obey the simple relation $m_2=0.11 (P_{\rm orb}/
1~{\rm hr})$ (e.g. Frank et al. 1992).  The prediction for $\dot m_T$
in case of GR driven mass transfer then becomes
\begin{equation}
\dot M_T \approx 8.85 \times 10^{14} m_1 (m_1+m_2)^{-1/3} \left
 ( P_{\rm orb} / 1 {\rm d} \right)^{-2/3} {\rm g~s^{-1}},~P_{\rm orb} <
 P_{\rm bif}~~~{\rm (GR)}.
\label{gr}
\end{equation}
which scales with the mass of the primary as $m_1^{2/3}$ for small
$m_2$. Note this is not very different from $\dot M_T \propto m_1$,
thus motivating the use of the Eddington scaling of luminosity in
Figs.~\ref{fig:lumtot} and~\ref{fig:lumfirm}

Equation~(\ref{ev}) for the case of secondary expansion is the linear
limit of third-order polynomial fits (King 1988) to detailed
evolutionary tracks of single stars computed by Webbink, Rappaport \&
Savonije (1983). Along an evolutionary track, the mass $m_2$ of the
expanding secondary decreases as it transfers mass onto the primary.
Since a variety of initial masses and orbital periods (at the onset of
mass-transfer) are possible, $m_2$ is essentially a free parameter in
equation~(\ref{ev}). However, a careful study of evolutionary
tracks\footnote{See Verbunt \& van den Heuvel (1995) for a
  semi-analytical model that allows one to compute tracks.} shows that
the band of mass transfer rates predicted by equation~(\ref{ev}) for
values of $m_2$ in the range 0.5 to 1 reproduces the mass transfer
rates in n-driven accreting binaries for the majority of their
lifetime. This band therefore provides, in a ``population synthesis''
sense, the likely mass transfer rates one can expect in SXTs with
orbital periods larger than the bifurcation period.

We estimate the mass transfer rate expected at a given orbital period
by simply adding the contributions from GR (Eq.~[\ref{gr}]) and
secondary expansion (a band with $m_2=[0.5,1]$ in Eq.~[\ref{ev}]).
Figure~\ref{fig:accrate} shows that this results in a relatively
short, although plausible, bifurcation period $\sim 10$ hr.  The
estimates of $\dot M_T$ are shown separately for BH and NS SXTs, in
units of $10^{18}$ g s$^{-1}$. The mass transfer rates are typically
similar, in absolute units, in BH and NS SXTs with long orbital
periods.  However, the mass transfer rate in BH SXTs is larger than in
NS SXTs at short orbital periods (cf Eq.~[\ref{gr}] and its scaling
with $m_1$); the two are in fact quite close in Eddington units at
these periods, where the majority of SXTs are found.  Although this
result neglects any contribution from magnetic braking, we think that
it is significant enough to motivate the use of Eddington-scaled
luminosities in Figs.~\ref{fig:lumtot} and~\ref{fig:lumfirm}. Note
that the comparable values of Eddington-scaled $\dot m_T$ predicted
for BH SXTs and NS SXTs at short periods contrast with the large
Eddington-scaled luminosity difference of the two classes of objects
(Figs.~\ref{fig:lumtot} and~\ref{fig:lumfirm}). This in essence is the
argument of Narayan et al. (1997b) and Garcia et al. (1998), and we
confirm their claim. The sample is, however, small.

Although the estimates shown in Fig.~\ref{fig:accrate} rely on simple
considerations that do not take into account the full complexity of
mass transfer (in particular, substantial fluctuations of the
mass transfer rates around the secular values used here may occur;
e.g. Warner 1995), they probably provide better than an
order-of-magnitude precision, which is sufficient for our
discussion. Note, however, that this simple model is not likely to
give an accurate value of $\dot M_T$ for the BH SXT J1655-40 since
this system seems to be in a special evolutionary state (crossing the
Hertzsprung gap, e.g. Kolb 1998).

\section{Accretion Models}

Following Narayan et al. (1996, 1997a), we assume that the accretion
flow is made of two components: an inner ADAF and an outer thin
accretion disk.  Since only the ADAF contributes to the emission of
quiescent SXTs in the 0.5-10 keV band, we ignore the thin
disk. Further, since the emission of the ADAF is primarily from
regions close to the center, we do not concern ourselves with the
precise value of the radius $R_{\rm tr}$ at which the transition
between the ADAF and the thin disk occurs.

Although the ADAF model was initially proposed for quiescent BH SXTs,
there is no reason why it should not apply also to quiescent NS SXTs.
In particular, if the transition of the accretion flow from a thin
disk configuration to an ADAF configuration is triggered by the same
local physics in the disk, e.g. a thermal instability in the upper
layers of the disk (Shaviv \& Wehrse 1986; Meyer \& Meyer-Hofmeister
1994; Narayan \& Yi 1995b), the process should occur independently of
the nature of the central object. We note, however, that the presence
of a reradiating surface in the case of a NS primary at the center of
the ADAF may have significant effects on the ADAF. For instance,
Narayan \& Yi (1995b) showed that the critical accretion rate $\dot
m_{\rm crit}$ in an ADAF is reduced in the case of a NS (compared to a
BH) because the additional supply of soft photons (from the stellar
surface) causes the ADAF to cool more efficiently. This is not an
issue here since $\dot m$ in all the quiescent systems we study is
well below $\dot m_{\rm crit}$.

We assume conservative mass transfer.  Consequently, if $\dot M_{\rm
ADAF}$ is the rate at which mass is accreted during quiescence via the
ADAF and $\dot M_{\rm accum}$ is the rate at which mass is accumulated
in the outer thin disk, the mass transfer rate satisfies $\dot
M_T=\dot M_{\rm ADAF} + \dot M_{\rm accum}$.  The proportion of mass
accreted vs. mass accumulated in quiescence is, however, uncertain.
Menou et al. (1998) estimated the relative importance of $\dot M_{\rm
ADAF}$ and $\dot M_{\rm accum}$ for most of the BH SXTs listed in
Table~\ref{tab:systems}. They estimated $\dot M_{\rm accum}$ by taking
the integrated fluences of the outbursts of each source and assuming a
radiative efficiency of $\eta = 0.1$ (which provides the total mass
accreted in the outburst), and estimated $\dot M_{\rm ADAF}$ by
fitting the quiescent emission with an ADAF model.  Menou et al.'s
work indicates that, in quiescent BH SXTs, roughly half the matter
that is transferred from the secondary is accreted via the ADAF and
the rest is accumulated in the thin disk, i.e.
\begin{equation}
\frac{\dot M_{\rm ADAF}}{\dot M_{\rm accum}} \equiv g \sim 1,~~~
\frac{\dot M_{\rm ADAF}}{\dot M_T} = \frac{g}{1+g} \sim \frac{1}{2}.
\label{eq:g}
\end{equation}
In the following models, we treat $g$ as a free parameter, but we
expect $g$ to be of order unity both in BH SXTs and in NS SXTs.

The version of the ADAF model used here is described in detail by
Narayan et al. (1998a). It consistently includes adiabatic compressive
heating of electrons in the energy equation (Nakamura et al. 1997) and
it uses for the flow dynamics the general relativistic global
solutions calculated by Popham \& Gammie (1998).

In previous studies involving ADAFs, most of the model parameters were
kept fixed at the following values (see Narayan et al. 1998b for a
discussion): $\alpha_{\rm ADAF}=0.3$ (the viscosity parameter in the
ADAF), $\beta \equiv P_{\rm gas} / P_{\rm total}=0.5$ (i.e.
equipartition between the gas and the magnetic field),
$\gamma=13/9=1.44$ (adiabatic index of the gas) and $\delta=10^{-3}$
(fraction of the viscous dissipation that goes directly into heating
the electrons).  We adopt these standard values of $\alpha_{\rm
  ADAF}$, $\beta$ and $\gamma$ in the present models (see \S 6 for the
effect of varying $\delta$). We take $\dot M_T$ to be given by the
predictions of binary evolution models (Fig.~\ref{fig:accrate}) and
obtain $\dot M_{\rm ADAF}$ via equation~(\ref{eq:g}) for a given
assumed value of $g$.  We calculate the spectral energy distributions
of the models and numerically integrate the spectra to obtain the
luminosities in the 0.5-10 keV band. The results are discussed in the
following sections.

\subsection{ADAF models of BH SXTs}

Figure~\ref{fig:lumbh}a shows two bands of luminosities corresponding
to ADAF models of BH SXTs. In the upper band $\dot M_{\rm ADAF} = \dot
M_T$, i.e.  the entire $\dot M_T$ is accreted by the black hole via
the ADAF, while in the lower band $\dot M_{\rm ADAF} = \dot M_T /3$
(or $g=1/2$). The lower band is in satisfying agreement with the
observed luminosities.  The fact that we obtain agreement with a
reasonable value of $g$ means that the modeling of the quiescent X-ray
emission of BH SXTs with the ADAF model is consistent with the
predictions of binary evolution models. The increase in the luminosity
with increasing $P_{\rm orb}$ is also in good agreement. One should
stress again, however, that the agreement for the BH SXT J1655-40 is
unexpected if the system is crossing the Hertzsprung gap (Kolb 1998).

The bolometric radiative efficiency in the ADAF models constructed
here is typically $\sim 10^{-3}$ compared to $\sim 10^{-1}$ for
standard thin disk accretion.  The X-ray radiative efficiency of these
ADAF models is another two or three orders of magnitude lower since
much of the ADAF luminosity comes out as synchrotron radiation in the
optical.  Note that these results are sensitive to various details of
the ADAF models, and in particular to the choice of $\alpha_{\rm
  ADAF}$ and $\beta$. For instance, we find that, if $\alpha_{\rm
  ADAF}=0.1$, we need $g \sim 1/4$ (i.e. a fraction $\sim 1/5$ of the
mass supplied by the secondary must be accreted via the ADAF) in order
to obtain agreement with the observed quiescent luminosities of BH
SXTs. The fraction goes down further to $\sim 1/10$ if $\alpha_{\rm
  ADAF}=0.025$.

\subsection{ADAF models of NS SXTs}

The presence of a hard surface on the accreting star is the key
feature of ADAF models of NS SXTs. The large amount of energy advected
in the ADAF, which is lost through the event horizon in BH SXTs, will
be reradiated from the NS surface (Narayan \& Yi 1995b). Narayan et
al. (1997b) and Garcia et al. (1998) argued that this additional
source of luminosity is the explanation for the systematically larger
luminosity of quiescent NS SXTs as compared to quiescent BH SXTs
(Figs~\ref{fig:lumtot} and \ref{fig:lumfirm}). We now investigate this
quantitatively.

In the following, we assume that the thermal energy stored in the ADAF
is reradiated as blackbody emission with a radiative efficiency $\eta
\sim 0.1$. The luminosity in the 0.5-10 keV band will be dominated by
the emission coming from the NS surface, so that the additional
emission from the ADAF can be ignored (for simplicity). We also assume
that the reradiation occurs from a small fraction $f_{\rm surf}$ of
the NS. This is in agreement with observations of quiescent NS SXTs
(e.g. Verbunt et al. 1994; Campana et al. 1998; but see Rutledge et
al. 1998 and the discussion in \S 7). For now, we do not seek to
justify the small emitting area of the NS. We simply assume for all
our NS models that the area is $10~{\rm km^2}$. However, our results
are not seriously affected even if the radiating area is increased or
decreased by, say, an order of magnitude (i.e as long as most of the
blackbody emission is in the 0.5-10 keV X-ray band). The typical
effective temperatures found are $T_{\rm eff} \lta 0.5$ keV.

The upper band in Fig.~\ref{fig:lumbh}b shows the luminosities
predicted by binary evolution models if $g=1/2$, i.e a third of the
mass transferred by the secondary is accreted onto the NS (as in
BH SXTs). We see that the luminosities predicted are much larger than
those observed, by $\sim 3$ orders of magnitude (except for the system
EXO 0748-676; see \S 7 for a discussion of this system). Since any
contribution to the mass transfer by magnetic braking would only
increase the mass transfer rates expected (in short orbital period
systems), the discrepancy between the observed and predicted
luminosities would be even more serious if magnetic braking were
effective.

One possibility for the discrepancy is that most of the quiescent flux
from NS SXTs is emitted outside the 0.5-10 keV X-ray band. This
explanation seems, however, rather unlikely since this emission would
be seen somewhere else in the spectrum, say in hard X-rays or soft
$\gamma$-rays. Aql X-1 has been observed in quiescence in hard X-rays
and it is clear that the energy does not come out in the 10-100 keV
band (Campana et al. 1998). In addition, although some theoretical
models of boundary layers between an accretion flow and a NS predict
substantial deviations from blackbody emission at higher energies
(e.g. Shapiro \& Salpeter 1975, Turolla et al. 1994), there are no
models predicting that such a large fraction ($\sim 0.999$) of the
emission would come out in hard X-rays or $\gamma$-rays.

We must therefore consider the possibility that only part of the mass
supplied by the secondary actually reaches the surface of the NS.  The
lower band in Fig.~\ref{fig:lumbh}b shows that the luminosities of
quiescent NS SXTs can be accounted for (except EXO 0748-676) if only
$\sim 0.1 \%$ of the transferred mass reaches the NS surface. This is
of course a surprisingly small fraction.\footnote{Although this value
  might simply be interpreted, in a scenario without ADAF, as the
  accretion rate at the inner edge of an unsteady quiescent disk
  ($\dot M \propto R^{2.5-3}$ in such a disk; e.g. Cannizzo 1993), we
  do not favor this interpretation because the propeller effect should
  operate in this case and prevent any mass from reaching the NS
  surface (see \S 5)}. In the following section, we argue that the
propeller effect can explain the fraction if combined with accretion
via an ADAF, while in \S 6 we consider the possibility that winds from
ADAFs could further reduce the amount of mass reaching the NS surface.

\section{Propeller Effect in Quiescent Neutron Star SXTs}

The propeller effect in accreting neutron stars was initially proposed
by Illarionov \& Sunyaev (1975) to explain the existence of
long-period X-ray pulsars, and it was later developed in more detail
by Davies and Pringle (1981; see also Wang \& Robertson 1985; Stella
et al. 1986). Although the details vary, the basic mechanism of
the effect relies on the presence of a centrifugal barrier at the
rotating magnetosphere of a rapidly spinning NS.

Inside the magnetosphere, the accreting gas is forced to follow the
magnetic field lines of the NS since, by definition, in this region
magnetic forces dominate the flow dynamics.  The fate of the gas is
then determined by the relative magnitudes of the magnetospheric
radius, $R_{\rm m}$, and the corotation radius, $R_{\rm co}$, defined
by $\Omega_{\star} \equiv (GM/R_{\rm co}^3)^{1/2}$, where
$\Omega_{\star}$ is the angular velocity of the NS (and the
magnetosphere).  If $R_{\rm co} < R_{\rm m}$, the magnetosphere
rotates so fast that the centrifugal force at $R_{\rm m}$ is larger
than the force of gravity and hardly any of the gas can be accreted
onto the NS.  This is the propeller effect. The efficiency of the
propeller effect is not well understood (see Davies, Fabian, \&
Pringle 1979, and Davies \& Pringle 1981).  It is also not clear
whether the accreting matter is ejected from the system or is merely
accumulated in a boundary layer around the magnetosphere (Wang \&
Robertson 1985).  However, from the simple physical argument outlined
above, it is clear that normal accretion onto the neutron star is
inhibited during the propeller phase.

The first direct observational evidence for the existence of the
propeller effect in accreting neutron stars was reported by Cui (1997)
in two X-ray pulsars, GX 1+4 and GRO J1744-28.  Cui showed that in
both systems X-ray pulsations ceased during a period of low X-ray
flux, which he interpreted as the result of a decrease in the mass
accretion rate and a corresponding increase in the magnetospheric
radius to beyond the corotation radius. Campana et al. (1998) and
Zhang et al. (1998) argued that the propeller effect was also seen in
a NS SXT, Aql X-1, during its most recent outburst; the system showed
an abrupt decrease of the X-ray flux, accompanied by considerable
hardening of the X-ray spectrum.  Both sets of authors proposed that
the hard spectrum originates just outside the magnetosphere, where the
gas becomes very hot due to the action of the propeller (see also Cui
et al. 1998).

Observations of Aql X-1 show that even in quiescence the blackbody
spectral component does not disappear entirely implying that a small
amount of material continues to accrete onto the star, contrary to
simple models of the propeller (Verbunt et al. 1994; Stella et al.
1994).  To resolve this problem, Zhang et al. (1998) proposed that
accretion in quiescent NS SXTs occurs via a quasi-spherical ADAF,
rather than a thin disk. This allows some material to accrete near the
poles, thereby bypassing the centrifugal barrier.  In this section we
develop this idea quantitatively.
 
\subsection{Accretion Geometry in Quiescence}

In the following, we assume for simplicity that the NS spin axis is
perpendicular to the binary orbital plane and that the NS spin and
magnetic axes are aligned.

\subsubsection{Magnetospheric Radius}

In the presence of a spherical accretion flow (e.g. Bondi 1952), the
magnetospheric (or Alfv\`en) radius, $R_{\rm m}$, is usually defined
as the radius at which the magnetic pressure, $P_{\rm mag} = B^2/8
\pi$, due to the neutron star magnetic field balances the ram
pressure, $P_{\rm ram} \equiv ~\rho v_r^2/2$, of the accreting gas,
where $v_r$ is the radial infall velocity of the gas (e.g. Frank et
al. 1992).  The contribution from the thermal pressure of the gas is
generally neglected when considering Bondi-type flows which are highly
supersonic (free-falling).  An ADAF, on the other hand, is subsonic
down to a few Schwarzschild radii (e.g. Narayan, Kato, \& Honma 1997;
Chen, Abramowicz \& Lasota 1997; Abramowicz, Chen, Granath \& Lasota
1996) and the thermal pressure term must be taken into account.

The polar structure of ADAFs was described by Narayan \& Yi (1995a) in
the self-similar approximation.  They showed that in the limit of
strong advection ($f_{\rm adv} \sim 1$), the gas density and
temperature (and therefore the thermal pressure of the gas) are
essentially independent of the polar angle $\theta$; i.e. these
quantities are nearly constant on spherical shells centered on the
accreting object.  On the other hand, the radial infall velocity is
smaller than the thermal sound speed in the equatorial plane and
decreases towards the poles.  Clearly then, the thermal pressure of
the accreting material is the dominant term in calculating the
magnetospheric radius. Since this pressure is constant with $\theta$,
the magnetosphere will be nearly spherical.

Despite the difference between a Bondi spherical accretion flow and an
ADAF, the value of $R_{\rm m}$ is roughly the same in the two cases.
In the former, essentially all the gravitational energy released
during accretion goes into bulk kinetic energy of the flow, while in
ADAFs, the energy is stored as thermal energy of the gas.  Thus, the
ram pressure in Bondi flows must be of the same order as the thermal
pressure in ADAFs.  The magnetospheric radius in quiescence then takes
the same form (Frank et al. 1992):
\begin{equation}
\label{eq:Rmq}
R_{\rm mq} = 6.45 \times 10^5 \left( \frac{\dot M}{\dot M_{\rm Edd}} \right)^{-2/7} m_1^{-3/7} B_8^{4/7} R_{\rm
NS,6}^{12/7}~{\rm cm},
\end{equation}
where $B_8$ is the NS surface magnetic field in units of $10^8$ G and
$R_{\rm NS,6}$ is the NS radius in units of $10^6$ cm. 

For the remainder of the paper, $R$ refers to the radius in physical
units and $r$ refers to the radius in Schwarzschild units.

\subsubsection{Accretion in the Propeller Phase}

The propeller effect in binary systems is generally discussed in the
context of thin disk accretion.  In this model, once the propeller
becomes effective, no matter is able to reach the NS surface because
the centrifugal acceleration acts equally against gravity for all the
matter located in the disk orbital plane.

In a spherical flow, however, the centrifugal acceleration at the
magnetospheric radius acting on a parcel of gas, accreting at a polar
angle $\theta$ from the spin axis, is equal to $A_c= \Omega_{\star}^2 R_{\rm
mq} \sin \theta$, where $\Omega_{\star}$ is the angular speed of the
magnetic field lines anchored in the NS.  The direction of this force
is perpendicular to the spin axis of the NS (and parallel to the
orbital plane), so that the component of $A_c$ along the radial
direction is simply $A_c \sin \theta = \Omega_{\star}^2 R_{\rm mq} \sin ^2
\theta$.  The maximum polar angle below which the gravitational
acceleration ($= R_{\rm mq} \Omega_K^2(R_{\rm mq})$) wins over the
centrifugal force is given by
\begin{equation}
\sin \theta_0 = \frac{\Omega_K(R_{\rm mq})}{\Omega_{\star}}.
\label{eq:sin_theta_0}
\end{equation}
For $\theta_0 \ll 1$ this simplifies to 
\begin{equation}
\label{eq:theta_0}
\theta_0 \simeq \frac{\Omega_K(R_{\rm mq})}{\Omega_{\star}}.
\end{equation}

Note that the residual force acting on all parcels of gas, even those
which do not overcome the centrifugal barrier, tends to direct them
toward the orbital plane.  In this simple formulation, we do not
consider possible complications due to field line orientation effects.

\subsubsection{Fraction of Mass Reaching the NS Surface}

Only matter accreting between $\theta = 0$ and
$\theta = \theta_0$ can overcome the centrifugal barrier and reach the 
surface of the NS.  The mass accretion rate onto the star is then
\begin{equation}
{\dot M_{\rm NS}} = - 2 \int_{0}^{\theta_0} 2 \pi R \sin \theta \rho v_r
R d \theta,
\label{eq:mdotrtheta}
\end{equation}
where $\rho = \rho (R,\theta) \simeq \rho (R)$ is the gas density and
$v_r= v_r (R,\theta) \simeq v_r(R) \sin^2 \theta$ is the radial infall
velocity at angle $\theta$ (positive outward), as given by the
self-similar solution of Narayan \& Yi (1995a).  The factor of $2$ in
equation (\ref{eq:mdotrtheta}) is introduced to account for the two
polar caps.  The fraction of the total mass accretion rate $\dot
M_{\rm ADAF}$ in the ADAF that reaches the NS surface is then
\begin{eqnarray}
f_{\rm acc} &\equiv& \frac{\dot M_{\rm NS}}{\dot M_{\rm ADAF}} \simeq
\frac{2 \int_0^{\theta_0} 2 \pi R^2 \sin \theta \rho (R) v_r(R) \sin^2
\theta d \theta}{2 \int_0^{\pi / 2} 2 \pi R^2 \sin \theta \rho(R)
v_r(R) \sin^2 \theta d \theta} = \frac{2 \int_0^{\theta_0} \sin^3
\theta d \theta}{2 \int_0^{\pi / 2} \sin^3 \theta d \theta} \nonumber
\\ &\simeq& \frac{3}{8} \theta_0^4 \simeq \frac{3}{8} \left(
\frac{\Omega_K(R_{\rm mq})}{\Omega_{\star}} \right)^4,
\label{eq:Facc}
\end{eqnarray}
where the last two steps have been calculated in the limit of small
$\theta_0$. For a neutron star with a spin period of $2$ ms, a
magnetic field strength of $3 \times 10^8$ G and accreting at $\dot
M_{\rm ADAF}= 10^{-3} \dot M_{\rm Edd}$, we find $f_{\rm acc} \simeq
0.75 \times 10^{-3}$. Note that if the radial velocity does not vary
as $\sin^2 \theta$ but is more nearly constant with $\theta$ (as in
Bondi accretion), then $f_{\rm acc}$ would be larger (e.g., if $v_r
(R,\theta)$ is independent of $\theta$, $f_{\rm acc} \propto
\theta_0^2$ only). On the other hand, if the accretion flow has a
toroidal morphology with empty funnels along the rotation axis,
$f_{\rm acc}$ would be much lower than the expression given in
equation~(\ref{eq:Facc}), perhaps even zero. There is thus
considerable uncertainty in the value of $f_{\rm acc}$.

\subsubsection{Fraction of the NS Surface Emitting Radiation}

We assume that the gas accreted during the propeller phase follows the
magnetic field lines down to the NS polar caps, where its kinetic and
thermal energy is converted to radiation.  Dipolar field lines satisfy
the parametric equation (e.g. Frank et al. 1992):
\begin{equation}
R=C \sin^2 \theta,
\label{eq:dipoline} 
\end{equation}
where $C$ is constant for a given field line.  Thus, the field lines
intersecting the magnetospheric radius at $\theta=\theta_0$, emerge
from the NS surface at a polar angle $\theta_S$ given by:
\begin{equation}
\frac{\sin^2 \theta_S}{\sin^2 \theta_0}=\frac{R_{\rm NS}}{R_{\rm mq}}.
\label{eq:thetaS}
\end{equation}
The accreting material reaches the NS surface at $\theta\le\theta_S$,
and therefore the fraction of the neutron star surface that
re-radiates the accreted energy is
\begin{equation}
f_{\rm surf} \equiv \frac{2 \int_0^{\theta_S} 2 \pi \sin \theta d
\theta}{2 \int_0^{\pi / 2} 2 \pi \sin \theta d \theta} \simeq
\frac{\theta_S^2}{2} \simeq \frac{\theta_0^2}{2} \frac{R_{\rm
NS}}{R_{\rm mq}},
\label{eq:Fsurf}
\end{equation}
where we have made use of the fact that $\theta_S < \theta_0 \ll 1$.
Note the direct dependence of $f_{\rm surf}$ on the NS magnetic field
strength $B$ through $R_{\rm mq}$. For a neutron star with a spin
period of $2$ ms, a magnetic field strength of $3 \times 10^8$ G and
accreting at $\dot M_{\rm ADAF}= 10^{-3} \dot M_{\rm Edd}$, we find
$f_{\rm surf} \simeq 3 \times 10^{-3}$.

\subsection{Equilibrium Spin Frequency}

Equations (\ref{eq:Facc}) and (\ref{eq:Fsurf}) demonstrate that in
quiescence the mass accretion rate onto a NS and the emitting area,
which determine the total luminosity and spectrum of the system,
depend strongly on the NS angular rotation speed $\Omega_{\star}$ and
the field strength.  Although $\Omega_{\star}$ is not known for most
NS SXTs, we discuss in this section how it can be estimated using our
knowledge of the accretion history of an SXT and of the interaction
between the magnetosphere of the NS and the accretion flow.

During the propeller phase, in quiescence, the NS ejects most of the
accreting gas that reaches its magnetosphere.  The ejected material
leaves the system with higher specific angular momentum than it had
coming in, which results in an effective spin-down of the NS.
However, NS SXTs experience outbursts during which the mass accretion
rate onto the NS is orders of magnitude higher than in quiescence.
High $\dot M$ causes the magnetosphere to shrink inside the corotation
radius so that all the accreting gas is able to reach the NS surface.
The mass and angular momentum of the accreting gas is added to the NS,
which causes it to spin-up.  A NS SXT that accretes for a fairly long
time is expected to reach an equilibrium spin frequency $\Omega_{\rm
  \star, eq}$ such that the spin up during outburst is balanced by the
spin down during quiescence.

At the onset of mass transfer, the spin period of the NS in an SXT is
likely to be different from the equilibrium value.  In standard Ghosh
\& Lamb (1979) type models (for steady disks), the timescale
$\tau_{\rm eq}$ on which a typical NS reaches the equilibrium spin
period is usually small ($\gta 10^5$ yr) compared to the lifetime of
the LMXB ($10^8-10^9$ yr; see, e.g., Henrichs 1983 for a review).
Although $\tau_{\rm eq}$ may be a little larger for transient
accretion, it is reasonable to assume that most of the neutron stars
in SXTs have rotation rates close to $\Omega_{\star,\rm eq}$. It is
straightforward to show that $\Omega_{\star}$ is basically unaffected
by a single outburst or a single quiescent phase, because the time
spent between two outbursts ($\sim$ a few years) is much shorter than
$\tau_{\rm eq}$.  Therefore, it is not necessary to consider the small
jitter in $\Omega_{\star}$ during the quiescence-outburst cycle. In
the following, we simply assume that once a NS SXT has reached spin
equilibrium, then $\Omega_{\star} = \Omega_{\star,\rm eq}$ both during
outburst and quiescence.

\subsubsection{Idealized Model for the Spin-up and Spin-down Torques}

To determine the equilibrium spin frequency of a NS in a transient
binary system we need a theoretical description of the interaction
between the NS and the accretion flow.  We begin with the outburst phase
during which accretion occurs via a thin disk and the torque exerted
by the accreting gas on the NS is relatively well understood.

Despite some modifications, current models describing the interaction
between a NS and a magnetically threaded thin accretion disk remain
essentially identical to the model initially proposed by Ghosh \& Lamb
(1979). In this model, the spin-up torque on the NS results from two
contributions. First, the gas that follows magnetic field lines and
reaches the NS surface gives its angular momentum to the NS (spin-up).
Second, the interaction between the magnetic field lines and the
threaded thin disk beyond the magnetospheric radius results in a
positive torque on the NS for radii where the field lines rotate
slower than the local Keplerian angular speed of the gas. The lines
that thread the disk further out give a (smaller) negative torque on
the NS since these field lines rotate faster than the local Keplerian
angular speed of the gas (Gosh \& Lamb 1979). Recent detailed
numerical simulations by Daumerie (1996) show that the overall
contribution resulting from the interaction of the accretion flow and
the magnetic field lines beyond the magnetospheric radius is nearly
equal to the contribution to the spin-up from the gas reaching the NS
surface. Further, Daumerie finds that the torque varies roughly
linearly with the fastness parameter $\omega \equiv
\Omega_{\star}/\Omega_K(R_{\rm m})$.

Using this work as a guide, we propose the following idealized
formula for the torque:
\begin{equation}
\dot{J}= 2 \dot{M} R_{\rm m}^2 \Omega_K (R_{\rm m}) \left[ 1- \left(
\frac{\Omega_{\star}}{\Omega_K (R_{\rm m})}\right)\right],
\label{eq:idealtorque}
\end{equation}
where the factor of 2 appears because of the two nearly equal
contributions to the spin-up mentioned above.
Equation~(\ref{eq:idealtorque}) assumes that the fastness parameter at
equilibrium, $\omega_{\rm crit}$, is exactly unity. This is consistent
with the recent arguments of Wang (1995; see also Wang 1987).

The torque in the propeller regime, when $\Omega_{\star} >
\Omega_K(R_{\rm m})$, is not as well constrained.  By analogy with
Ghosh \& Lamb type models, we assume that the torque arises from two
nearly equal contributions. One contribution is the negative torque
applied to the NS when the propeller expels gas at the magnetosphere
and the other is the negative torque\footnote{This contribution to the
torque is always negative in the propeller regime because
$\Omega_{\star} > \Omega_K(R)$ for all $R > R_{\rm mq}$.} due to the
interaction of the magnetic field lines with the accretion flow beyond
the magnetospheric radius. If the gas expelled leaves the
magnetosphere with an angular momentum corresponding to the angular
rotation speed of the NS, viz. $\Omega_{\star}$ (which corresponds to
an efficient propeller), then the negative torque on the NS in the
propeller phase is
\begin{equation}
\dot{J} \sim - 2 \dot{M} R_{\rm m}^2 \Omega_{\star}, 
\label{eq:torquepropel}
\end{equation} 
where $\dot M$ is the rate at which mass reaches the magnetosphere.
(The small positive contribution from the very small fraction of mass
accreted on to the NS is neglected.)  This expression is in agreement
with equation~(\ref{eq:idealtorque}) in the limit of $\Omega_{\star}
\gg \Omega_K (R_{\rm m})$.  Therefore, we can use the prescription
given by equation~(\ref{eq:idealtorque}) to describe the interaction
between the NS and the accretion flow throughout the
outburst--quiescence cycle.

\subsubsection{Spin Equilibrium}

We define $\Delta J_o=\dot J_o \Delta t_o$ to be the integrated
spin-up torque over an outburst of duration $\Delta t_o$, and $\Delta
J_q=\dot J_q \Delta t_q$ to be the integrated spin-down torque during
a propeller phase of duration $\Delta t_q$. The recurrence time of the
SXT is $\Delta t_o + \Delta t_q$.  We assume, for simplicity, that the
two phases can be approximated as bimodal, with the outburst phase
occurring at a typical accretion rate $\dot M_o$ and the quiescent
phase having a typical lower accretion rate $\dot M_q$. These
accretion rates define, in turn, the values of the magnetospheric
radius in outburst,\footnote{Following Frank et al. (1992), we use a
value for the magnetospheric radius of the thin disk that is half that
used for spherical accretion (Eq.~[\ref{eq:Rmq}])} $R_{\rm mo}$, and
in quiescence, $R_{\rm mq}$, used in equation~(\ref{eq:idealtorque})
to determine $\dot J_o$ and $\dot J_q$. Since the accretion rate from
the peak of the outburst is known to decrease gradually with time, the
value of $\dot M_o$ used to determine $R_{\rm mo}$ should be
interpreted as a mean over the outburst.

By definition, the NS angular rotation speed at equilibrium
$\Omega_{\star,\rm eq}$ satisfies $\Delta J_o=\Delta J_q$. Using
equation~(\ref{eq:idealtorque}), we find
\begin{equation}
\Omega_{\star,\rm eq}= \Omega_K (R_{\rm mq}) \left( \frac{R_{\rm
mq}}{R_{\rm mo}} \right)^{3/2} \frac{\left[ 1+ \left( \frac{R_{\rm
mq}}{R_{\rm mo}} \right)^{1/2} \tilde{g}\right]}{\left[ 1+ \left( \frac{R_{\rm
mq}}{R_{\rm mo}} \right)^{2} \tilde{g} \right]},
\label{eq:eqomega}
\end{equation}
where $\tilde{g}$ is the ratio $\dot M_q \Delta t_q / \dot M_o \Delta
t_o$, i.e. the total mass reaching the magnetosphere during the
ADAF-propeller phase divided by the total mass accreted during
outburst. This definition of $\tilde{g}$ is equivalent to the
definition of $g$ given in equation~(\ref{eq:g}) if all the mass
accumulated in the disk during quiescence is accreted onto the neutron
star during outburst.\footnote{A possible situation where the mass
  accreted onto the neutron star during outburst is less than the mass
  accumulated in the disk during quiescence is if the depletion rate
  of the disk during outburst is super-Eddington while the accretion
  rate onto the neutron star is Eddington-limited.  Even in that case,
  Eq.~(\ref{eq:eqomega}) remains valid because $\Omega_{\star,\rm eq}$
  depends on the ratio $\dot M_q \Delta t_q / \dot M_o \Delta t_o$}
For the remainder of the paper, we assume $g=\tilde{g}$.  Note that
the spin equilibrium period defined by equation~(\ref{eq:eqomega})
differs from the spin equilibrium period defined for {\it steady}
accretion onto a NS and usually discussed in the literature (e.g.
Henrichs 1983).

\subsection{Propeller Regime at Spin Equilibrium}

Equation~(\ref{eq:eqomega}) allows us to determine the equilibrium
spin period $P_{\rm spin}$ as a function of the parameter $g=\dot
M_{\rm ADAF}/\dot M_{\rm accum}$. This, in turn, allows us to estimate
the quantities $f_{\rm acc}$ (Eq.~[\ref{eq:Facc}]) and $f_{\rm surf}$
(Eq.~[\ref{eq:Fsurf}]) which are relevant for determining the
observational properties of quiescent NS SXTs. In the following
calculations, we assume for definiteness that the NS accretes at the
Eddington rate ($\dot M_o=\dot M_{\rm Edd}$) in outburst and that the
mass transfer accretion rate in quiescence is $\dot M_T=10^{-2.5} \dot
M_{\rm Edd}$ (cf Fig.~\ref{fig:accrate}). However, our conclusions do
not depend crucially on these assumptions.

Figures~\ref{fig:eqfacc} and~\ref{fig:eqbyp} show the variations of
$\dot M_{\rm NS}/ \dot M_T= f_{\rm acc} \times g/(1+g)$, $f_{\rm
  surf}$ and $P_{\rm spin}$ with $g$. As $\dot M_{\rm ADAF}$
decreases, the magnetosphere becomes more extended and the propeller
effect becomes more efficient. The solid lines show the results for a
NS with a magnetic field strength $B=10^8$ G, while the dashed and
dotted lines show the cases $B=10^9$ and $B=10^{10}$ G, respectively.
Recent observations suggest that accreting neutron stars in low mass
binaries have rather low magnetic fields, typically $B \lta 10^9$ G
(e.g. White \& Zhang 1997).

Figure~\ref{fig:eqfacc} can be related to Fig.~\ref{fig:lumbh}b, where
we found that $\dot M_{\rm NS}/ \dot M_T \sim 10^{-3}$ was needed in
order to explain the very low observed luminosities of quiescent NS
SXTs. Such a value of $\dot M_{\rm NS}/ \dot M_T$ is predicted by the
propeller model if $g \sim 1/5$. This value of $g$ is reasonably close
to the value $g \sim 1/2$ that we obtained for BH SXTs.  Note that the
dependence of $\dot M_{\rm NS}/ \dot M_T$ on the magnetic field $B$
should cancel out through the ratios $R_{\rm mq}/R_{\rm mo}$ in
equation~(\ref{eq:eqomega}). For low magnetic field strengths,
however, the inner edge of the disk in outburst is not given by half
the value in equation~(\ref{eq:Rmq}) but is fixed at the last stable
orbit at 3 Schwarzschild radii (for instance, if $\dot M_o \geq 0.1
\dot M_{\rm Edd}$ and $B \leq 10^9$ G). Consequently, there is a
residual dependence of $\dot M_{\rm NS}/ \dot M_T$ on $B$ for small
$B$.

Figure~\ref{fig:eqbyp}a shows that the fraction $f_{\rm surf}$ of the
NS surface that emits in quiescence depends much more strongly on $B$
than $f_{\rm acc}$ does (Fig.~\ref{fig:eqfacc}), as expected from
equation~(\ref{eq:Fsurf}). For values of $g \sim 1/5$ and low magnetic
fields ($B \lta 10^9$ G), the emitting surface is typically between
$1$ and $10~{\rm km^2}$. Observations of quiescent NS SXTs suggest an
emitting surface $\sim 1~{\rm km^2}$ (e.g. Verbunt et al. 1994;
Campana et al. 1998; but see \S 7). Given that the surface could very
well be underestimated by the standard techniques (e.g. Lewin, Van
Paradijs \& Taam 1993; Rajagopal \& Romani 1996; Zavlin et al. 1996;
Rutledge et al. 1998), values of $\sim 1-10~{\rm km^2}$ appear in
reasonable agreement with the observations. We note, however, that
this result depends quite crucially on the assumed alignment of the NS
spin and magnetic axes.  This is a major uncertainty in the
calculations.

Figure~\ref{fig:eqbyp}b shows that $P_{\rm spin}$ is even more
sensitive to the value of $B$. For weak magnetic fields ($B \lta
10^9$ G), the values of the equilibrium spin period are in the range
$1-10$ ms, which is in agreement with values recently discussed in the
literature for accreting neutron stars in low mass binaries
(e.g. White \& Zhang 1997).

There are several sources of uncertainty in our simplified model for
the propeller, which are discussed in \S 6.2 and \S 7 below.  The
proximity of the value $g \sim 1/2$ required to explain the quiescent
luminosities of BH SXTs to the value $g \sim 1/5$ required to explain
the luminosities of NS SXTs (with a propeller) suggests, however, that
roughly the same geometry of accretion (outer thin disk and inner ADAF
sharing the mass supplied by the secondary roughly equally) can
account for the properties of both classes of objects. The presence of
a hard surface and a magnetic field associated with the spinning
neutron star are two major ingredients of the model. The NS SXTs Aql
X-1 ($B \sim 10^8$ G, $P_{\rm spin} \sim 2-3$ ms, emitting area $\sim
1$ km$^2$; White \& Zhang 1997; Zhang et al. 1998; Campana et al.
1998) and SAX J1808.4-3658 ($B < 2 \times 10^8$ G, $P_{\rm spin} =
2.49$ ms; Wijnands \& van der Klis 1998; Chakrabarty \& Morgan 1998,
hereafter CM98) are two interesting candidates that could allow
further testing of the predictions of our ADAF-propeller scenario.

We note that CM98 inferred a mass transfer rate in SAX J1808.4-3658
$\approx 10^{-11}$ M$_{\odot}$ yr$^{-1}$, which is roughly a factor
ten less than the value for a 2 hr orbital period NS SXT according to
our Fig.~\ref{fig:accrate}. These authors find that the estimate is
consistent with mass transfer being driven by angular momentum losses
due to gravitational radiation, if the secondary star has a mass
$m_2=0.05$. The values shown in Fig.~\ref{fig:accrate}, however, were
deduced assuming that the donor star is a main sequence star which
satisfies $m_2=0.11(P_{\rm orb}/{\rm 1~hr})$ (\S 3). This is
apparently not the case in SAX J1808.4-3658, perhaps because of the
fact that the companion star is strongly irradiated (CM98).

\section{Winds from ADAFs and the Efficiency of the Propeller Effect}

The gas in an ADAF could in principle escape to infinity via a wind
because part of the flow may be unbound. Narayan \& Yi (1994, 1995a)
showed that the Bernoulli constant, a measure of the energy of the gas
at infinity, is positive in their self-similar solution, and Narayan,
Kato \& Honma (1997) showed that this is also the case in a global
solution. A simple way of taking this into account was proposed by
Blandford \& Begelman (1998; hereafter BB98) who derived self-similar
ADAF solutions that include mass loss via winds.

Our results in \S 4.1 suggest that reducing $\alpha_{\rm ADAF}$ has
only a small effect on the radiative efficiency of the accretion flow.
Quataert \& Narayan (1998) showed, however, that an efficient way of
increasing the radiative efficiency of the ADAF is to include winds
and simultaneously increase the value of the parameter $\delta$, the
fraction of the viscously dissipated energy that directly heats the
electrons. They found that models with winds and high $\delta$ are
consistent with the current observations for the BH SXT V404 Cyg in
quiescence and the Galactic Center source Sgr A*.  Note that
theoretical estimates of the value of $\delta$ are highly uncertain.
Quataert (1998) and Gruzinov (1998) argued that $\delta$ could be
small if the magnetic fields in the ADAF are fairly subthermal, but
Bisnovatyi-Kogan \& Lovelace(1997; see also Quataert \& Gruzinov 1998)
argued that magnetic reconnection in the accretion flow could
preferentially heat the electrons and lead to large values of $\delta$
($\sim 1$).

In \S 6.1 and \S 6.2, we investigate the following two questions. Can
winds by themselves, i.e. without any propeller effect, explain the
observed quiescent luminosities of NS SXTs?  If not, is a combination
of winds and a propeller consistent with the observations?

\subsection{ADAF Models of NS and BH SXTs with Winds}

Following BB98, we assume that the accretion rate in the ADAF scales
in a self-similar way, i.e.
\begin{equation}
\dot m(r)=\dot m(r_{\rm tr})\left(\frac{r}{r_{\rm tr}}\right)^p,
\label{eq:Rmqwind}
\end{equation}
where $\dot m$ is the accretion rate scaled in Eddington units ($\dot
m=\dot M/\dot M_{\rm Edd}$), $0<p<1$, and we assume that the wind is
effective over the entire radial extent of the ADAF out to $r_{\rm
  out}=r_{\rm tr}$, the transition radius.

The efficiency of mass loss is not only determined by $p$, but also by
the radial extent of the ADAF (Eq.~[\ref{eq:Rmqwind}]). Around a black
hole, the ADAF extends from the transition radius ($r_{\rm tr}$) down
to the event horizon, and so the accretion rate onto the black hole is
$\dot m_{\rm in}=\dot m(r_{\rm tr})(1/r_{\rm tr})^p$. Around a
neutron star, however, the ADAF has a minimum radius equal to the
magnetospheric radius $r_{\rm m}$, so that the accretion rate at the
magnetosphere is $\dot m_{\rm in}=\dot m(r_{\rm tr})(r_{\rm m}/r_{\rm
  tr})^p$.  If $r_{\rm m} \gg 1$, as is the case for $r_{\rm mq}$ in
quiescent NS SXTs, then for a given value of $p$, the presence of a
magnetosphere around the NS significantly reduces the total mass loss
relative to the BH case. The value of $r_{\rm m}$ itself depends on
the local accretion rate at the magnetosphere (Eq.~[\ref{eq:Rmq}]).
Since the accretion rate at a given radius decreases with increasing
$p$, a larger $p$ implies both a larger $r_{\rm m}$ and a smaller
radial extent for the ADAF.

We solve for the magnetospheric radius by combining
equations~(\ref{eq:Rmq}) and~(\ref{eq:Rmqwind}) and find
\begin{equation}
r_{\rm mq}=\left[1.35 \times \dot m(r_{\rm tr})^{-2} r_{\rm
tr}^{2p} B_8^4 R_{\rm NS,6}^{12} \right]^{\frac{1}{7+2p}},
\label{eq:Rmqwind2}
\end{equation}
where $r_{\rm mq}$ is expressed in Schwarzschild units for a NS of 1.4
$M_\odot$ (Eq~[\ref{eq:Rmqwind2}] is the generalization of
Eq~[\ref{eq:Rmq}] for non-zero $p$). We have confirmed (by combining
Eqs.~[\ref{eq:Rmqwind}] and~[\ref{eq:Rmqwind2}]) that despite the
reduction of the radial extent of the ADAF for larger values of $p$,
increasing $p$ leads to an effective reduction of $\dot m(r_{\rm mq})$
for any reasonable values of $p$, $r_{\rm tr}$ and $\dot m(r_{\rm
  tr})$.

For a standard NS with low magnetic field strength ($B_8=1$, $R_{\rm
  NS,6}=1$) and a typical accretion rate at the outer boundary of the
ADAF $\dot m(r_{\rm tr})=1/3 \times 10^{-2.5}$ (\S 4.2;
Fig.~\ref{fig:accrate}), we find $r_{\rm mq} \simeq 10$ in the absence
of a wind ($p=0$). For a wind with $p=0.4$, $r_{\rm tr}=10^4$
(``intermediate'' wind model, $\delta = 0.3$), we find $r_{\rm mq}
\simeq 20$ and $\dot m(r_{\rm mq})/\dot m(r_{\rm tr}) \simeq 8 \times
10^{-2}$, while for a wind with $p=0.8$, $r_{\rm tr}=10^4$ (``strong''
wind model, $\delta = 0.75$), we find $r_{\rm mq} \simeq 35$ and $\dot
m(r_{\rm mq})/\dot m(r_{\rm tr}) \simeq 1 \times 10^{-2}$.  For these
estimates, we have chosen $r_{\rm tr} =10^4$ by analogy with the
values usually inferred in quiescent BH SXTs (Narayan et al. 1996;
Menou et al. 1998). Were $r_{\rm tr}$ to be smaller, the overall mass
loss in the wind for a given $p$ would be smaller. Note that the ADAF
cannot extend far beyond $10^4$ Schwarzschild radii because, for a
given $\dot m (r_{\rm tr})$, there is a maximum radius out to which
the ADAF can exist (e.g. Esin et al.  1997; Menou et al. 1998).

Following Quataert \& Narayan (1998), we computed two examples of ADAF
models assuming that the dominant effect of the wind on the structure
and the emission properties of the ADAF is to reduce the density in
the accretion flow; in addition, we assumed that the emission from the
wind itself can be neglected relative to the emission from the
accretion flow.

Figure~\ref{fig:bhwind}a shows the quiescent X-ray luminosities (0.5-10
keV) of BH SXTs predicted by the ADAF+wind model for which $\dot
m(r_{\rm tr})$ is equal to $1/3$ of the mass supplied by the
secondary, $p=0.4$, $r_{\rm tr}=10^4$ and $\delta=0.3$.  This model
gives roughly the same X-ray luminosities as the previous no-wind ADAF
model (\S 4.1). The actual mass accreting onto the black hole is
reduced here, but nevertheless the luminosities are the same because
of the increase in the value of $\delta$ from $10^{-3}$ to $0.3$ (see
Quataert \& Narayan 1998).

Figure~\ref{fig:bhwind}b shows the quiescent X-ray luminosities (0.5-10
keV) of NS SXTs predicted by the same ADAF+wind model, assuming that
all the mass reaching the magnetosphere is accreted onto the NS
surface (i.e. no propeller effect). Since the X-ray luminosity of the
ADAF is still much less than the luminosity coming from the NS
surface, the main effect of including the wind is to decrease the
luminosity of quiescent NS SXTs in proportion to the reduction of
$\dot m$ at the inner edge of the ADAF (Eq.~[\ref{eq:Rmqwind}]).  By
comparing figures~\ref{fig:lumbh}b and~\ref{fig:bhwind}b, it is clear
that the luminosity is smaller by roughly an order of magnitude, which
is insufficient to explain the observations.  Thus, the intermediate
wind scenario must also invoke a propeller effect, though with a
somewhat lower efficiency than that described in \S5.  Given the
uncertainties in our propeller model (see \S6.2), we see no objection
to such a model.

Only for the strongest wind models ($p \sim 1$, $r_{\rm tr}=10^4$)
does the luminosity difference between NS SXTs and BH SXTs predicted
by the models reduce to the observed difference. For $p=1$ and $r_{\rm
tr}=10^4$ ($B_8=1$, $R_{\rm NS,6}=1$), Eq.~(\ref{eq:Rmqwind})
and~(\ref{eq:Rmqwind2}) predict $r_{\rm mq} \simeq 58$ and $\dot
m(r_{\rm mq})/\dot m(r_{\rm tr}) \simeq 6 \times 10^{-3}$, which is
basically consistent with the small accretion rates required to
explain the observed quiescent luminosities of NS SXTs
(cf. Fig.~\ref{fig:lumbh}b). But is this a reasonable model? Can 
the propeller effect be completely ineffective at preventing mass
from reaching the NS surface?

\subsection{The Efficiency of the Propeller Effect}

If the neutron stars in NS SXTs have spin periods of typically a few
ms, as suggested by recent observations (e.g. White \& Zhang 1997;
Zhang et al. 1998; Campana et al. 1998; Wijnands \& van der Klis
1998), their corotation radii are a few Schwarzschild radii.  This is
substantially smaller than the few tens of Schwarzschild radii
inferred for $r_{\rm mq}$ in \S 6.1. The propeller effect is therefore
expected to act quite strongly in these systems (Illarionov \& Sunyaev
1975). Note that the values of $r_{\rm mq}$ quoted in \S 6.1, which
correspond to a magnetic field strength $B$ of $10^8$ G, would be even
larger for a larger $B$ (eq.~[\ref{eq:Rmqwind2}]).  Can the propeller
in such a system be so ineffective that $f_{\rm acc} \sim 1$?

We pointed out in \S5.2 that there are large uncertainties in the
values of $f_{\rm acc}$, which come mainly from the strong dependence
of $f_{\rm acc}$ on the NS rotation speed and the magnetospheric
radius: $f_{\rm acc} \propto \Omega_{\star}^{-4} R_{\rm mq}^{-6}$ (the
dependence on $R_{\rm mq}$ also affects the size of the NS emitting
area, since $f_{\rm surf} \propto R_{\rm mq}^{-4}$).  The latter
source of uncertainty is perhaps more important, since while
$\Omega_{\star}$ can be determined by direct observations, $R_{\rm
  mq}$ must be deduced from theoretical arguments.

Although the value of $R_{\rm mq}$ used in our work
(Eq.~[\ref{eq:Rmq}]) is standard, it is based on simple dimensional
arguments. Given the strong dependence of $f_{\rm acc}$ on $R_{\rm
  mq}$, an error, for instance, of only a factor of $1.5$ in $R_{\rm
  mq}$ translates to an error of more than an order of magnitude in
$f_{\rm acc}$. Recent work by Psaltis \& Chakrabarty (1998) shows that
the standard scaling used for $R_{\rm mq}$ (for disk-magnetosphere
interactions) is not consistent with observations in at least one
weakly magnetized NS SXT (SAX J1808.4-3658). Consequently, we cannot
rule out the possibility that equation~(\ref{eq:Rmq}) overestimates
the value of $R_{\rm mq}$ for an ADAF-magnetosphere interaction. The
propeller effect would then be significantly less efficient than our
previous results suggest (\S 5).

If winds alone ($p \sim 1$) are to explain the reduced luminosity
difference between BH and NS SXTs in quiescence, we need a highly
inefficient propeller.  For example, from the no-wind ($p=0$) model to
the $p=1$ model described in \S 6.1, $r_{\rm mq}$ increases by a
factor $\simeq 5.8$ which implies a $\sim 4 \times 10^4$ times more
efficient propeller according to equation~(\ref{eq:Facc}). Since the
$p=0$ model has a propeller with $f_{\rm acc} = 10^{-3}$, the
propeller would have to be $\sim 10^{7}$ times less efficient in the
$p=1$ model than we would infer based on the theory described in \S 5.
In addition, in the presence of a wind, the value of $f_{\rm acc}$ is
likely only to be smaller than the expression in
equation~(\ref{eq:Facc}). This is because the wind flows out
preferentially from the polar regions of the ADAF (Narayan \& Yi
1995a), so there is no gas to flow down to the polar caps of the NS.
Even allowing for reasonable uncertainties in the scalings in
equation~(\ref{eq:Rmq}), it does not seem likely that the propeller
model could be that wrong. We therefore feel that the $p \sim 1$
strong wind model is unlikely.

\section{Discussion}

Chen et al. (1998; their Fig.~3) argued that the segregation in
luminosity swing between outburst and quiescence previously found by
Narayan et al. (1997b) and Garcia et al. (1998) disappears in a large
sample. We point out in this paper that the Chen et al. sample is
larger only because they include numerous upper limits, many of which
are based on observations with insufficient sensitivity to detect
typical quiescent SXTs. Unless treated with care, for instance with a
``detection and bounds'' type method (e.g. Avni et al. 1980; Schmitt
1985), upper limits can obscure a true correlation, and indeed this
seems to be the case with the analysis of Chen et al. (1998).

Our Fig.~\ref{fig:lumfirm} shows a carefully selected and smaller
sample, and we find that there is a difference of luminosity between
quiescent BH and NS SXTs.  Furthermore, we argue, based on theoretical
mass transfer rates, that it is important to compare quiescent
luminosities of NS SXTs and BH SXTs of similar orbital periods. We
find that quiescent BH SXTs with short orbital periods are about $2$
orders of magnitude fainter (in Eddington units) than quiescent NS
SXTs with comparable orbital periods.  The difference reduces to a
little more than one order of magnitude when absolute luminosities are
considered.  The sample is, however, small. Future observations with
{\it Chandra X-ray Observatory (CXO)} and {\it XMM} will hopefully
increase the sample size sufficiently to provide a more convincing
case. Hopefully, they will also allow separate comparisons of long and
short period systems.  This should help solidify the arguments
advanced by Narayan et al.  (1997b) and Garcia et al. (1998) for event
horizons in black holes.

We find that the luminosities of quiescent NS SXTs are lower than the
values predicted by standard binary evolution models when combined
with ADAF models. Following Zhang et al. (1998), we argue that the
propeller effect offers a plausible explanation for the discrepancy.
However, although our model for the propeller effect accounts for the
luminosities and small emitting areas of quiescent NS SXTs, it suffers
from several sources of uncertainty that may affect the results
significantly.

The predictions of the model strongly depend on the spin frequency
$\Omega_{\star}$ of the NS and therefore on the prescription chosen
for the torque (Eq.~[\ref{eq:idealtorque}]). The magnitude of the
torque is not well known, especially in the propeller regime (e.g.
Davies, Fabian \& Pringle 1979; Henrichs 1983). The results of our
calculations are also sensitive to the precise values of the
magnetospheric radii in quiescence and outburst and to the assumption
that the fastness parameter $\omega_{\rm crit}=1$.

If $\omega_{\rm crit}< 1$, $\Omega_{\star,\rm eq}$ is reduced by a
factor $\omega_{\rm crit}$ (Eq.~[\ref{eq:eqomega}]) and more mass is
accreted because $f_{\rm acc}$ is $\propto \omega_{\rm crit}^{-4}$.
Note, however, that independently of equation~[\ref{eq:idealtorque}]
and the value of $\omega_{\rm crit}$, spin periods of a few
milliseconds and low magnetic field strengths (assuming that
Eq.~[\ref{eq:Rmq}] is correct) lead to predictions for luminosities
and surface areas that are consistent with observations.

Even if the spin periods of neutron stars in NS SXTs are known, the
predictions of our propeller model are still somewhat uncertain
because of the strong dependence of the fractions $f_{\rm acc}$ and
$f_{\rm surf}$ on the magnetospheric radius in quiescence $R_{\rm
  mq}$. If the standard expression (Eq.~[\ref{eq:Rmq}]) overestimates
the value of $R_{\rm mq}$, then the propeller effect could be less
efficient than indicated by our results in \S5. In that case, winds
from ADAFs could provide the additional mechanism required to explain
the low quiescent luminosities of NS SXTs.

We have also neglected the likely misalignment between the neutron
star spin axis and magnetic axis and have assumed a dipolar structure
for the magnetic field. Relaxing these assumptions could also modify
our results, especially our estimates of the fractions $f_{\rm acc}$
and $f_{\rm surf}$.

The system EXO 0748-676 appears unusually bright in our sample of
quiescent NS SXTs. One possible explanation is that magnetic braking
acts more efficiently in this system because of its rather short
$P_{\rm orb}$, giving a larger mass transfer rate. If the mass
transfer rate is higher, the propeller effect would also be less
efficient, and would increase the mass falling on the NS even further.
Note that EXO 0748-676 is known to be an unusual binary system, as
shown by its unexplained variations of $P_{\rm orb}$ (Hertz, Wood \&
Cominsky 1997). Another possible explanation is that, for some reason
(youth?), this particular system has not yet reached equilibrium and
spins at a relatively slower rate than other NS SXTs. This would result
in a less efficient propeller and a more luminous system in
quiescence.

This last point is a general prediction of our propeller model: the
propeller effect is less efficient if the NS in a SXT spins slowly
($P_{\rm spin} \gg$ a few ms). It is possible that several such NS
SXTs with slowly spinning NSs exist in our Galaxy. In these NS SXTs,
most of the mass accreted via the ADAF would reach the NS in
quiescence. The outbursts of these systems would therefore be of small
amplitude, 2-3 orders of magnitude in X-rays rather than 5-6 orders of
magnitude, because of their relatively high quiescent luminosities.

In our model, the gas which is stopped by the propeller is neglected
and its fate is left unspecified. This gas could affect both the
dynamics and the emission properties of the accretion flow. For
instance, it is unclear how the mass accreted via the ADAF will find
its way out of the quasi-spherical accretion flow after being
propelled outward.

In \S 6, we considered ADAF models with winds assuming that the
efficiency $p$ of the wind is the same for BH and NS SXTs . In
principle, $p$ could be different for ADAFs around BHs and around NSs
(because of the presence of the magnetosphere in NSs for instance).
Having a different value of $p$ for BH and NS SXTs would not, however,
affect our conclusion that winds alone probably cannot explain the
observed quiescent luminosities of NS SXTs.

Recently, Brown et al. (1998) argued that nuclear reactions in the
crust of neutron stars, triggered during outbursts in NS SXTs, could
efficiently heat up the NS cores in these systems.  The energy
deposited during an outburst would be reemitted during quiescence at a
rate sufficient to explain the observed quiescent thermal emission.
If so, the propeller effect must be even more efficient than our
estimates in \S5 indicate, so that accretion in quiescence does not
contribute much to the quiescent emission of NS SXTs.

We note two criteria that bear on the relative importance of crustal
heating and accretion to the quiescent luminosity of NS SXTs.  First,
in the scenario of Brown et al., the entire surface of the NS
contributes to the quiescent emission.  Small emitting areas have
usually been inferred from the observations, by assuming that the
quiescent emission is close to blackbody emission. However, Rutledge
et al. (1998) have shown that much larger areas are inferred if
detailed hydrogen atmosphere models are used instead of idealized
blackbody models.  There is a large uncertainty in these results
because of the poorly known hydrogen column density to the systems.
Nevertheless, the values of the emitting areas inferred by Rutledge et
al for several systems are consistent with the entire NS surface being
responsible for the emission in quiescence.  This success favors the
crustal heating model.

On the other hand, any observed rapid variability (e.g. on timescales
of a few days) should be considered as strong evidence of the
importance of accretion in quiescence, because the intrinsic thermal
emission from the NS is not expected to vary on such small timescales
(Brown et al. 1998).  Campana et al. (1997) reported that the
quiescent X-ray luminosity of the prototypical NS SXT Cen X-4 varied
by a factor of $\sim 3$ over a period of $4$ days.This establishes
that the accretion luminosity can be more than 3 times larger than any
steady quiescent luminosity possibly present in this system.  Though
very little X-ray variability data are presently available, several
studies of the optical counterpart of Cen X-4 in quiescence have shown
a variability of order several tenths of a magnitude (e.g.  Cowley et
al.  1988; Chevalier et al.  1989; McClintock \& Remillard 1990).
Since it is natural to relate this optical variability to the X-ray
variability observed by Campana et al.  (1997), both sets of
observations appear consistent with accretion being the dominant
mechanism for the quiescent luminosity of NS SXTs.

Throughout this study, we assumed that accretion is responsible for
the quiescent emission of both NS and BH SXTs. The possibility that
this emission could be due to two different processes in NS SXTs
certainly complicates the comparisons between the two classes of
systems.  We note, however, that even if the dominant contribution to
the quiescent emission of NS SXTs is crustal heating, the reason why
quiescent NS SXTs, as a class, are brighter than quiescent BH SXTs
remains the presence of a hard surface in NS SXTs and of an event
horizon in BH SXTs.

\section{Conclusion}

In this paper, we have reconsidered the luminosity difference between
black hole and neutron star soft X-ray transients in quiescence, which
has been used to argue for the presence of event horizons in black
holes.  

We show that the current observational data suggest that quiescent BH
SXTs as a class are fainter than NS SXTs.  This result agrees with the
previous work of Narayan et al.  (1997b) and Garcia et al. (1998), but
it disagrees with the conclusions of Chen et al. (1998) who used an
inappropriate sample of SXTs for their comparison.

We point out that, for a reliable comparison of the two classes of
SXTs, objects with similar orbital periods $P_{\rm orb}$ should be
compared. Otherwise, variations of the mass transfer rate $\dot M_T$
with $P_{\rm orb}$ can mask the results.

We find that the observed luminosities of quiescent BH SXTs are
consistent with the predictions of binary evolution models for $\dot
M_T$ if roughly a third of the mass supplied by the secondary is
accreted by the black hole via an ADAF. This estimate is for
$\alpha_{\rm ADAF}=0.3$. The fraction goes down to $1/5$ and $1/10$
for $\alpha_{\rm ADAF}=0.1$ and $0.025$ respectively.  The observed
luminosities of quiescent NS SXTs suggest, on the other hand, that
only a very small fraction of the mass transferred by the secondary
reaches the neutron star surface.

We explain the small fraction in NS SXTs by invoking an efficient
propeller.  We have constructed a model for the propeller effect that
accounts for the observed luminosities of quiescent NS SXTs and their
small emitting areas. In addition, the model appears to be consistent
with the millisecond spin periods recently inferred from observations
of accreting neutron stars in transient low mass binary systems (Aql
X-1, SAX J1808.4-3658).

Winds from ADAFs constitute an alternative explanation (BB98) for the
very small fraction of mass reaching the neutron star surface in
quiescence. We argue that an ADAF model with strong winds and no
propeller cannot explain the observed luminosities.  However, an ADAF
model with a wind of low or intermediate strength and a somewhat less
efficient propeller (but still within the range of uncertainties of
our propeller model) is consistent with the observed quiescent
luminosities of BH and NS SXTs. In this case, the parameter $\delta$,
which measures the fraction of the viscous energy that goes directly
into the electrons, has to be large $\sim 0.3$ (Quataert \& Narayan
1998).

After submitting our paper for publication, we became aware of an
independent paper by Asai et al. (1998) on the same subject. The
conclusions of Asai et al. are very similar to those presented here:
(1) observational data show evidence for a difference of X-ray
luminosity between quiescent BH and NS SXTs, (2) only $0.1-1 \%$ of
the mass transferred by the secondary reaches the NS surface in
quiescent NS SXTs, and (3) this small accreted fraction could be due
to the propeller effect in NS SXTs.

The quiescent fluxes and upper limits reported herein are generally in
very good agreement with those listed in Asai et al. (1998), with the
exception of 4U1543-47.  Here we assume a distance of 8~kpc, while
Asai et al. assume 1~kpc.  This accounts for most, but not all of the
discrepancy between the Asai et al.  upper limit and that in Orosz et
al. 1998.  We choose to use the latter limit, as it is more
conservative.

\section*{Acknowledgments}

We are grateful to Josh Grindlay, Mario Livio, Philip Podsiadlowski,
Dimitrios Psaltis and Eliot Quataert for useful discussions, and an
anonymous referee for constructive comments.  AE and RN also
gratefully acknowledge their conversation with Shuang-Nan Zhang who
suggested the idea of a propeller effect in the context of a
quasispherical accretion flow. This work was supported in part by NASA
grant NAG 5-2837.  KM was supported by a Smithsonian Astrophysical
Observatory (SAO) Predoctoral Fellowship and a French Higher Education
Ministry Grant. AE was supported by a National Science Foundation
Graduate Research Fellowship and by NASA through AXAF Fellowship grant
PF8-10002 awarded by the AXAF Science Center, which is operated by the
SAO for NASA under contract NAS8-39073.

\clearpage

\begin{figure} 
\plotone{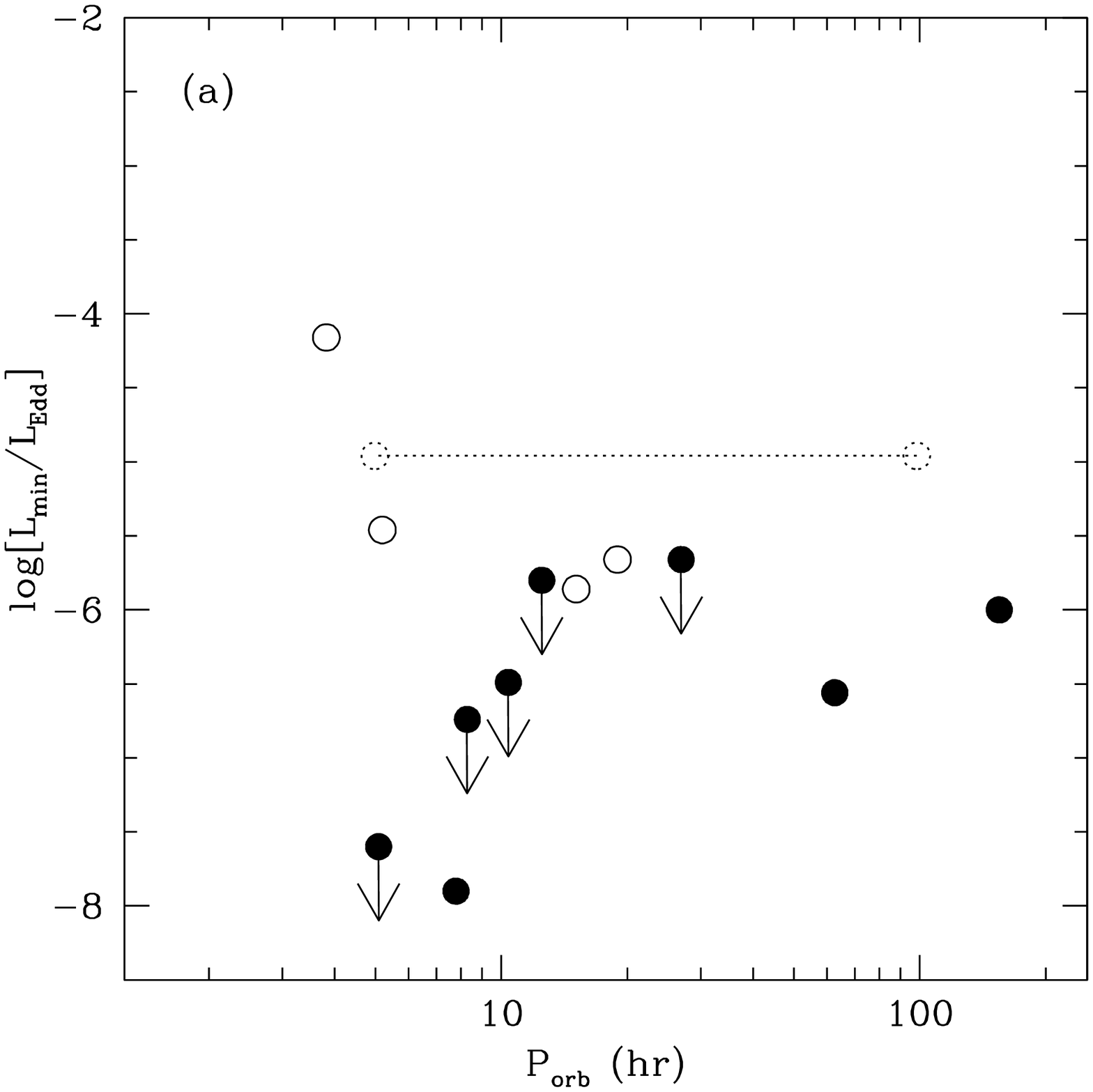}
\end{figure}
\begin{figure}
\plotone{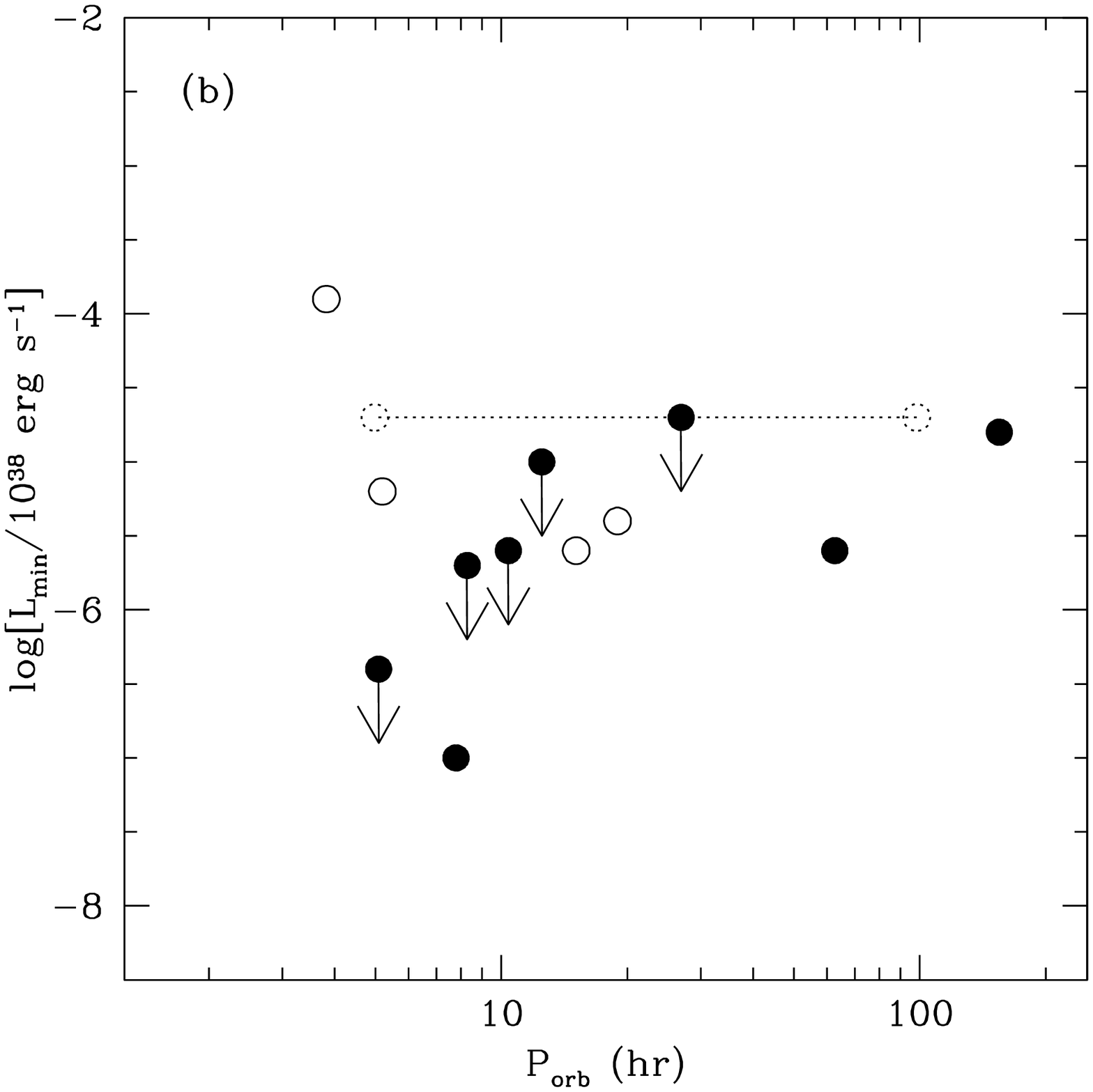}
\end{figure}
\begin{figure}
\plotone{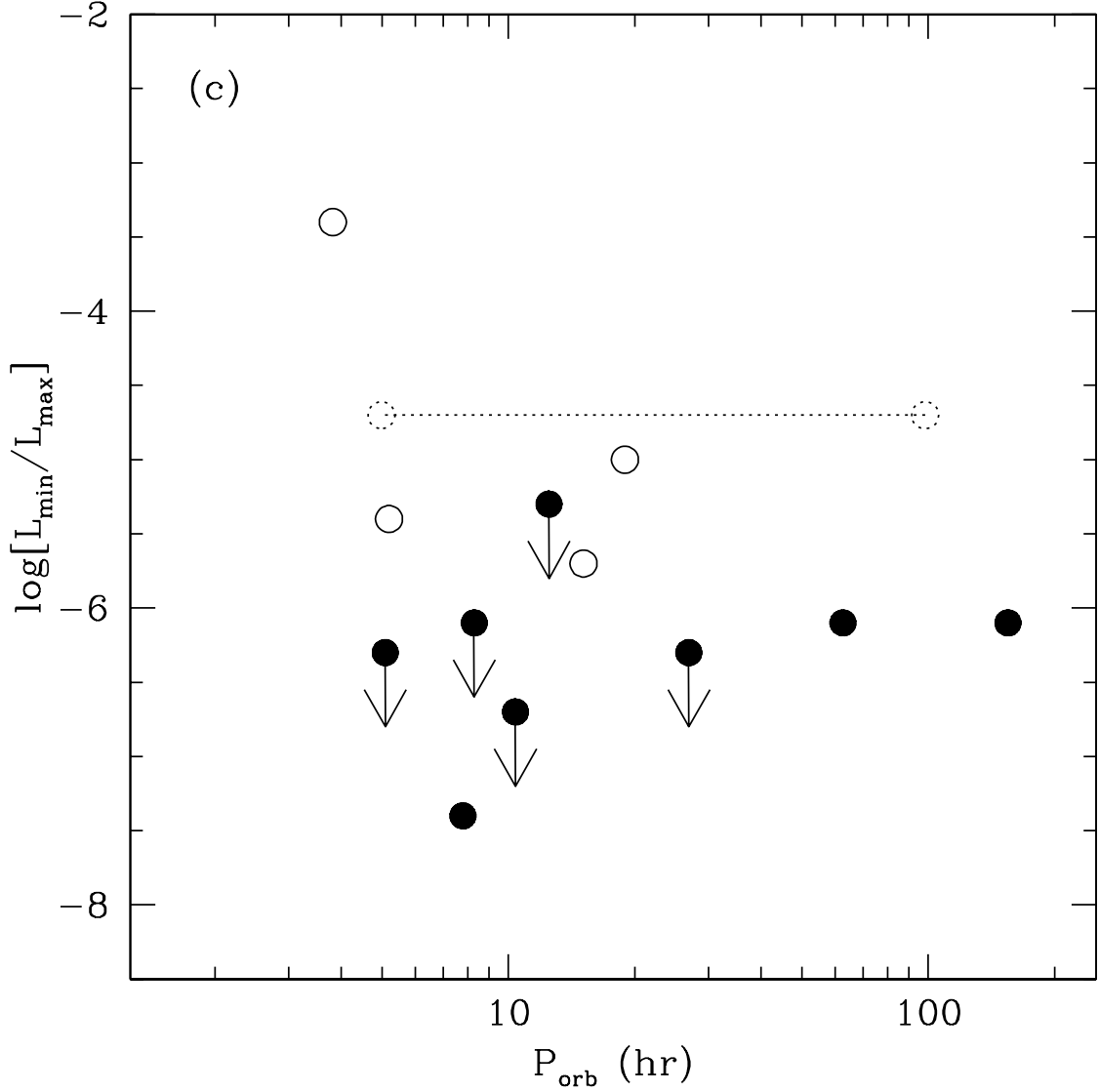}
\caption{(a) Quiescent luminosities $L_{\rm min}$ (Eddington units)
  in the 0.5-10 keV band of the NS SXTs (open circles) and BH SXTs
  (dots) listed in Table~1. The luminosities are plotted as a function
  of the orbital period $P_{\rm orb}$ of each system. (b) Same as (a)
  except that $L_{\rm min}$ are shown in units of $10^{38}$ erg
  s$^{-1}$. (c) Same as (a) except that the ratio $L_{\rm min}/L_{\rm
    max}$ is shown.
  \label{fig:lumtot}}
\end{figure}

\begin{figure} 
\plotone{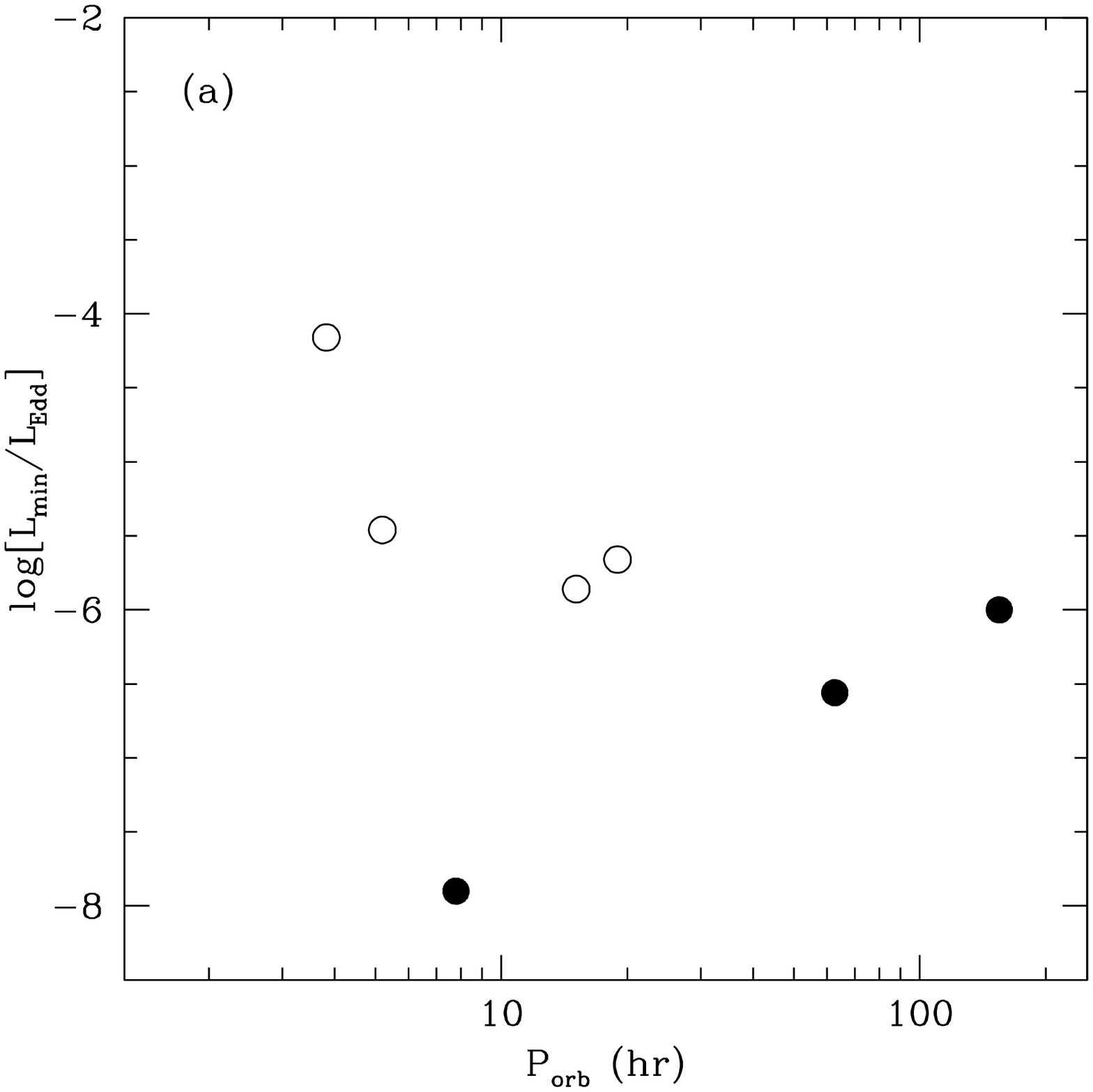}
\end{figure}
\begin{figure}
\plotone{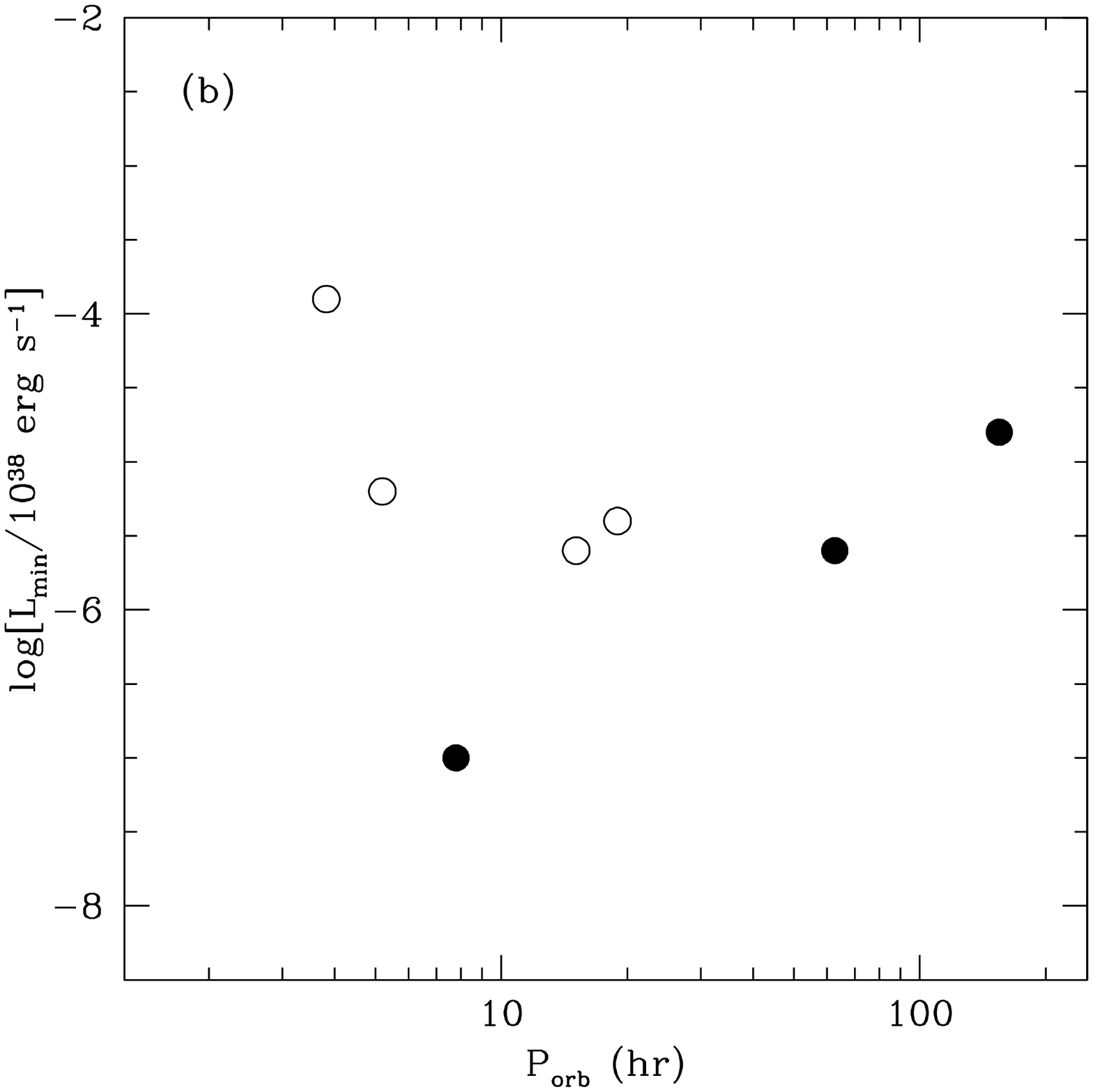}
\end{figure}
\begin{figure}
\plotone{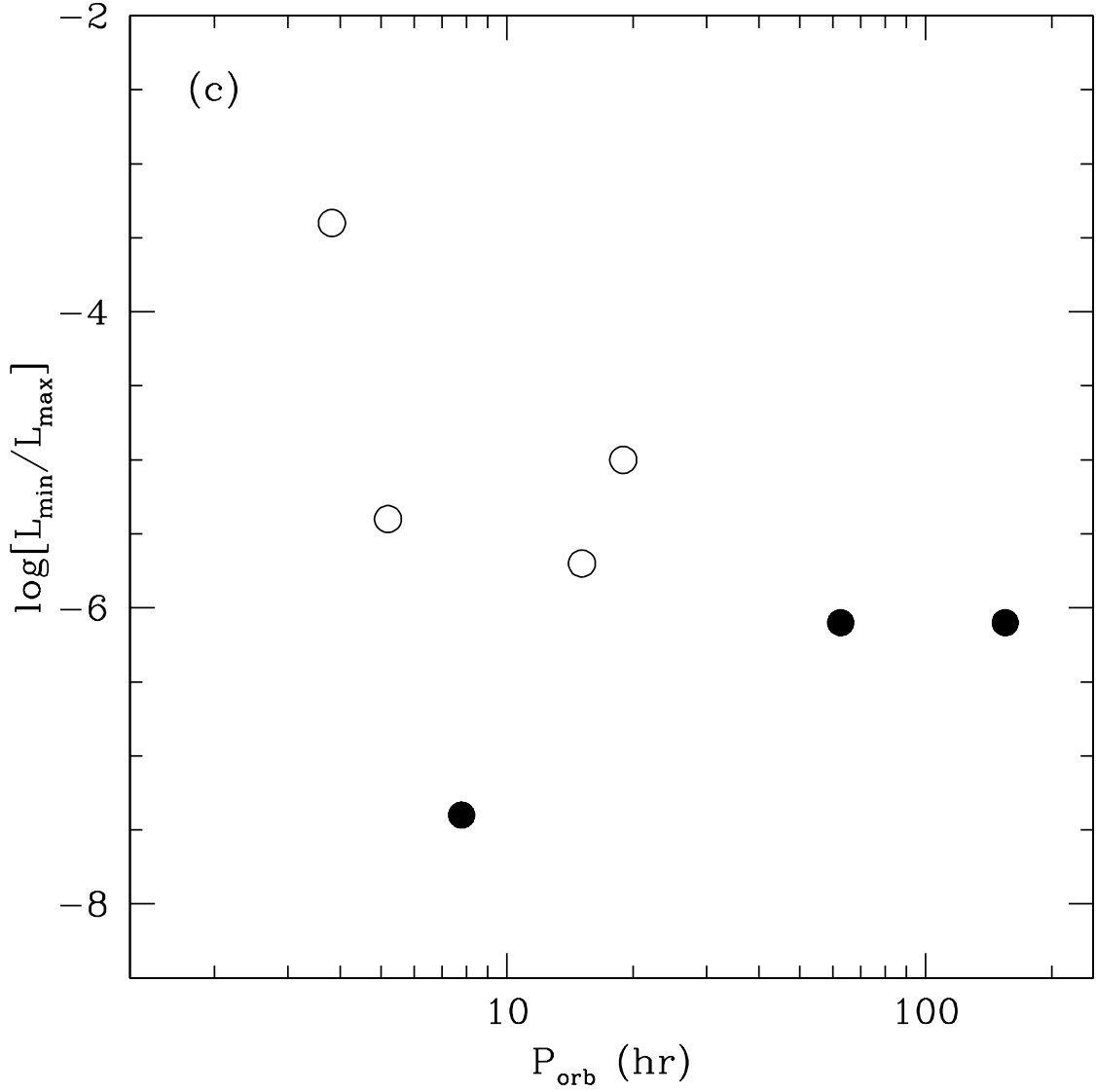}
\caption{(a) Same as Fig~1a, except that only SXTs with well
  determined $L_{\rm min}$ and $P_{\rm orb}$ are shown. (b) Same as
  (a) except that $L_{\rm min}$ are shown in units of $10^{38}$ erg
  s$^{-1}$. (c) Same as (a) except that the ratio $L_{\rm min}/L_{\rm
    max}$ is shown.\label{fig:lumfirm}}
\end{figure}

\begin{figure} 
\plotone{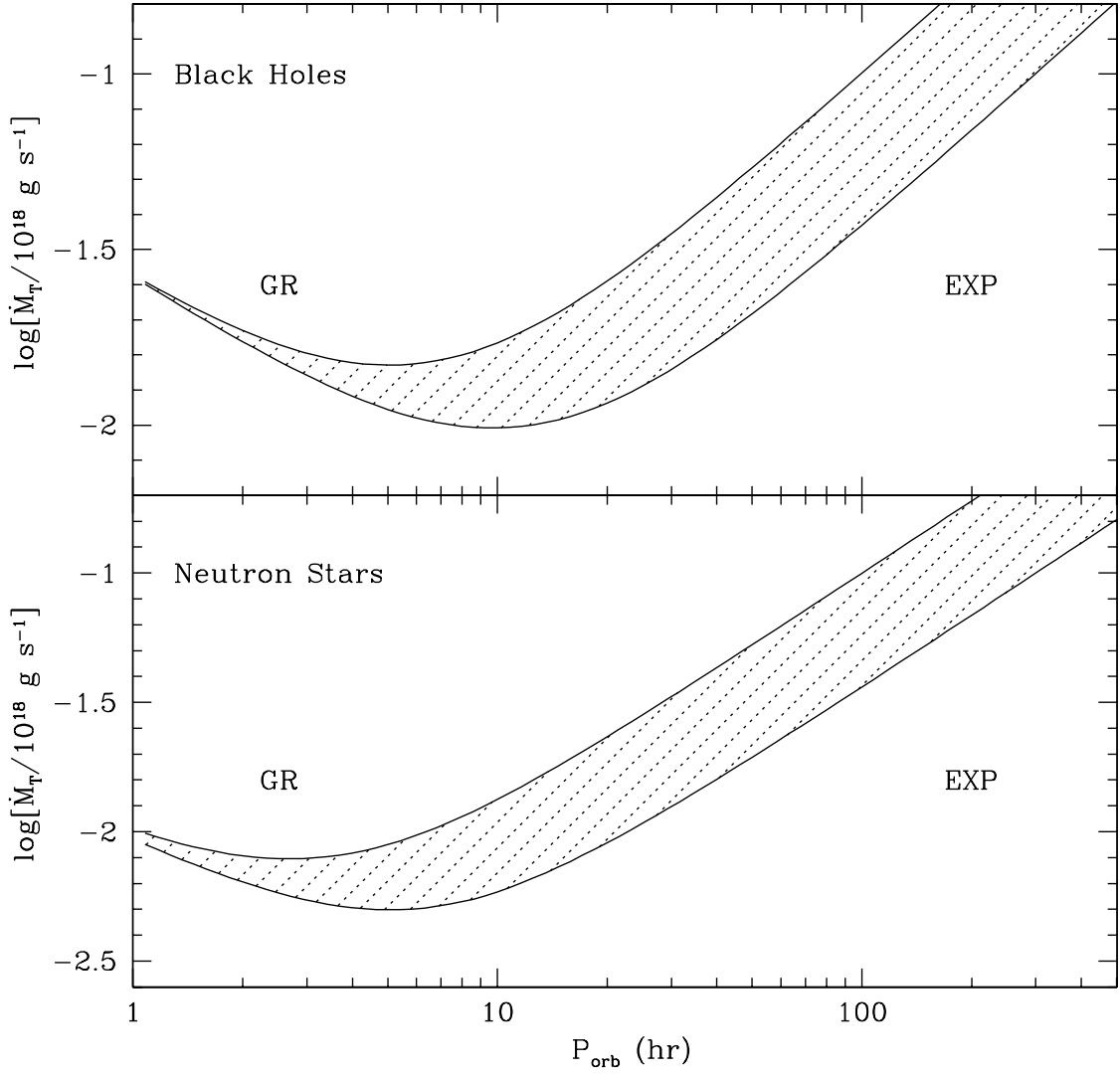}
\caption{Predictions of binary evolution models for the mass
  transfer rate $\dot M_T$ (in units of $10^{18}$ g s$^{-1}$) in NS and
  BH SXTs, as a function of the orbital period $P_{\rm orb}$.  GR
  refers to a phase of mass transfer driven by gravitational radiation
  and EXP to mass transfer driven by secondary
  expansion.\label{fig:accrate}}
\end{figure}

\begin{figure} 
\plottwo{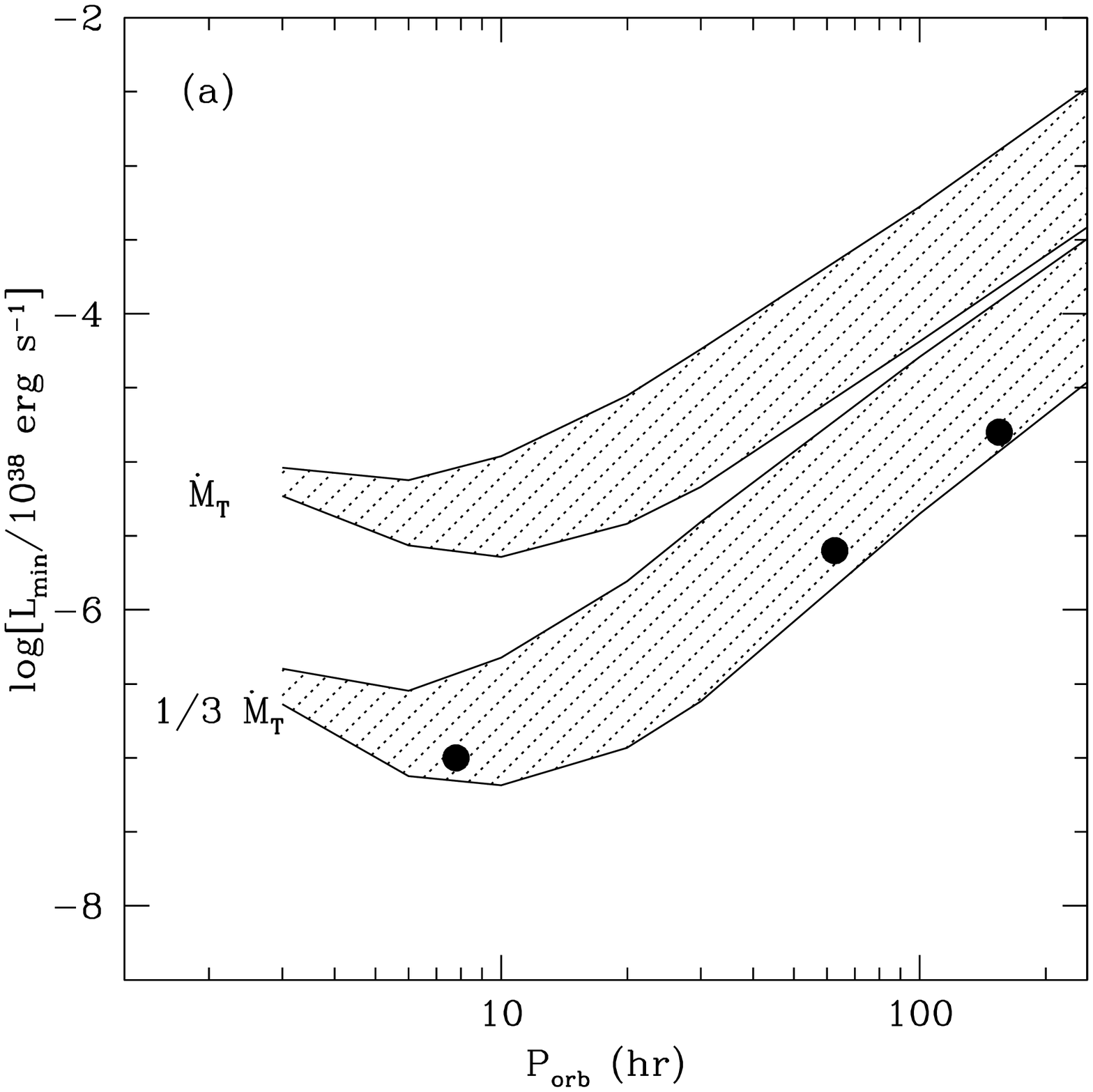}{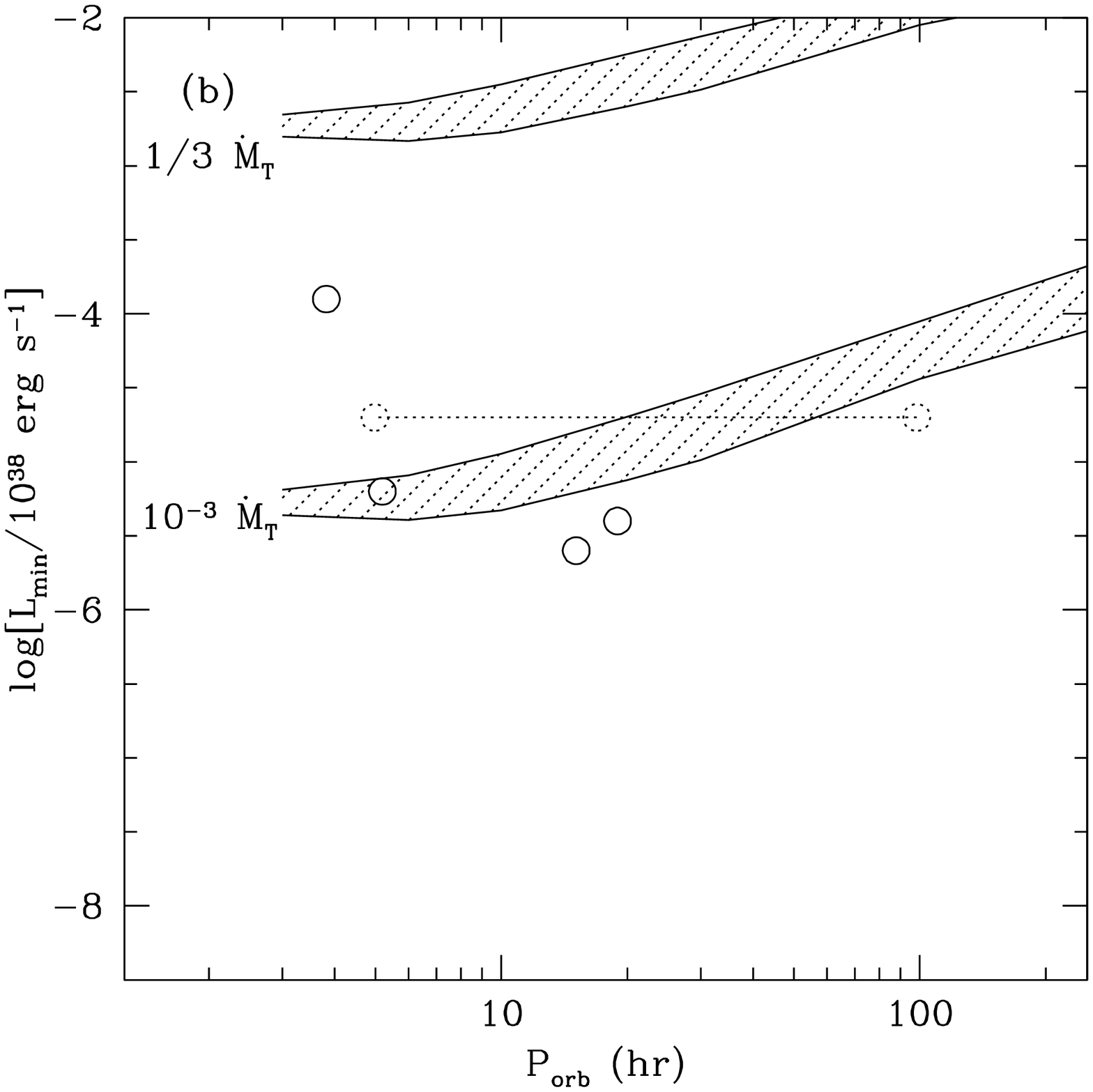}
\caption{(a) The upper band shows the quiescent luminosities of BH
  SXTs in the 0.5-10 keV band predicted by ADAF models if all the mass
  transferred by the secondary is accreted via the ADAF. The lower
  band corresponds to $\sim 1/3$ of the mass transferred being
  accreted via the ADAF.  This model fits the observed luminosities
  reasonably well. Both bands have been calculated for $\alpha_{\rm
    ADAF} =0.3$. The effect of varying $\alpha_{\rm ADAF}$ is
  explained in the text. (b) The upper band shows the quiescent
  luminosities of NS SXTs predicted in the 0.5-10 keV band if 1/3 of
  the mass transferred by the secondary reaches the NS surface. The
  lower band shows that the luminosities actually observed correspond
  to a very small fraction ($\sim 10^{-3}$) of the transferred mass
  reaching the NS surface.\label{fig:lumbh}}
\end{figure}

\begin{figure} 
\plotone{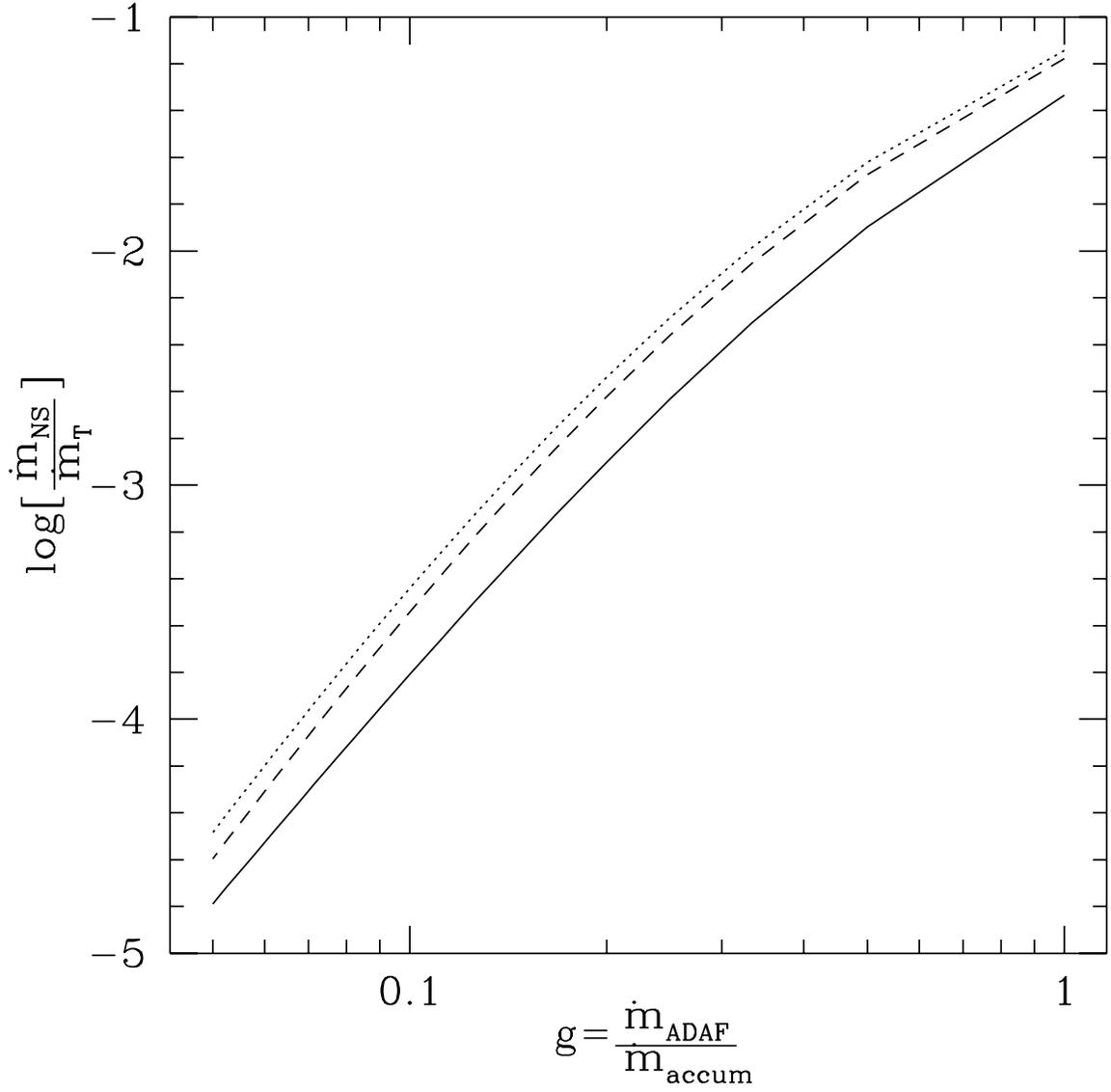}
\caption{Fraction of mass transferred by the secondary that reaches
  the neutron star surface in the propeller regime, as a function of
  $g=\dot M_{\rm ADAF}/\dot M_{\rm accum}$. The solid line shows this
  fraction for a neutron star with a surface magnetic field strength
  of $10^8$ G, while the dashed and dotted lines correspond to $10^9$
  G and $10^{10}$ G, respectively.\label{fig:eqfacc}}
\end{figure}

\begin{figure} 
\plottwo{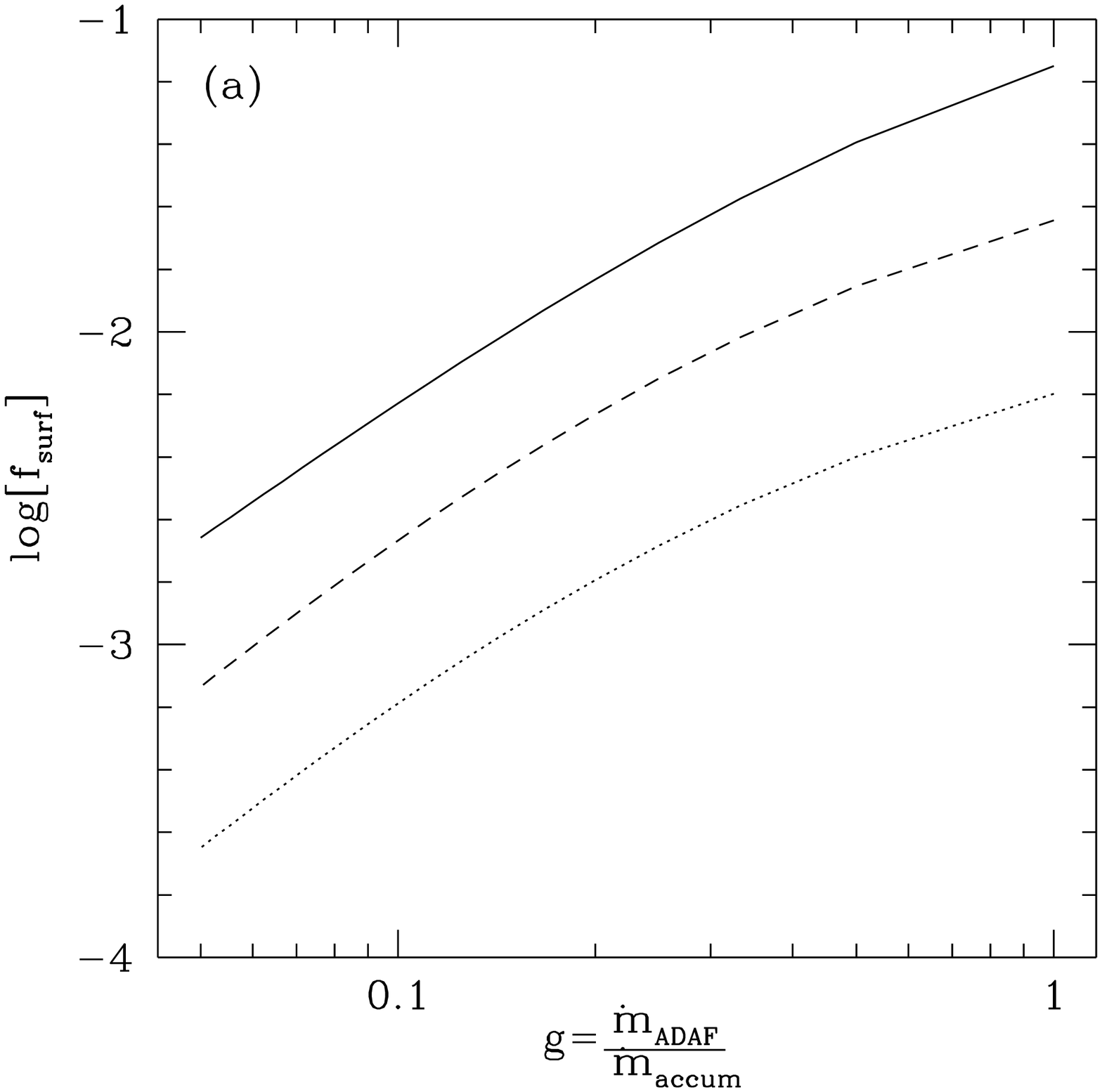}{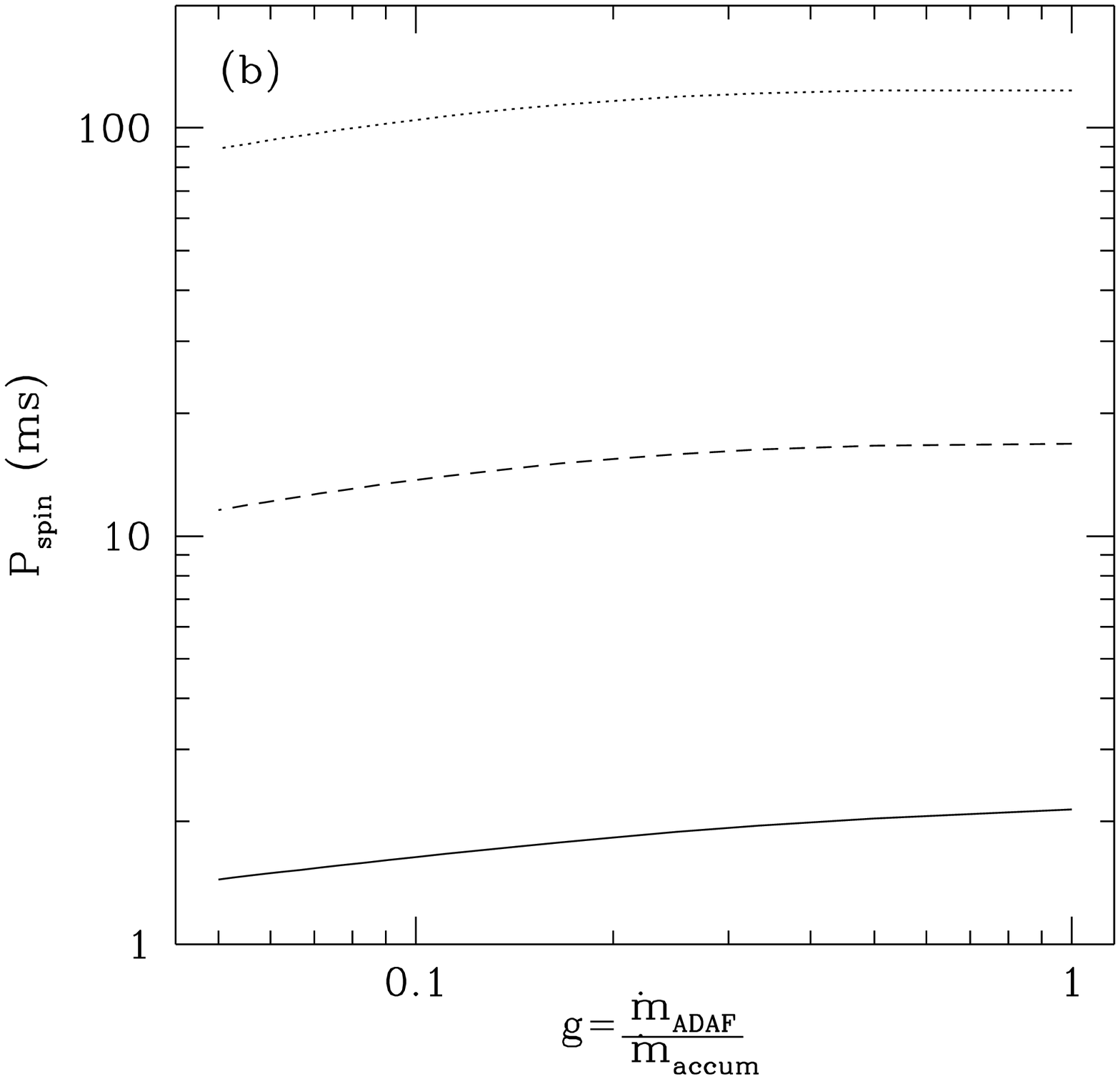}
\caption{(a) Fraction $f_{\rm surf}$ of the neutron star surface
  emitting radiation in the propeller regime, as a function of $g=\dot
  M_{\rm ADAF}/\dot M_{\rm accum}$. The solid line shows $f_{\rm
    surf}$ for a neutron star with a surface magnetic field strength
  of $10^8$ G, while the dashed and dotted lines correspond to $10^9$
  G and $10^{10}$ G, respectively. (b) Equilibrium spin period $P_{\rm
    spin}$ as a function of $g=\dot M_{\rm ADAF}/\dot M_{\rm accum}$
  for the same magnetic field strengths as in (a).\label{fig:eqbyp}}
\end{figure}

\begin{figure} 
\plottwo{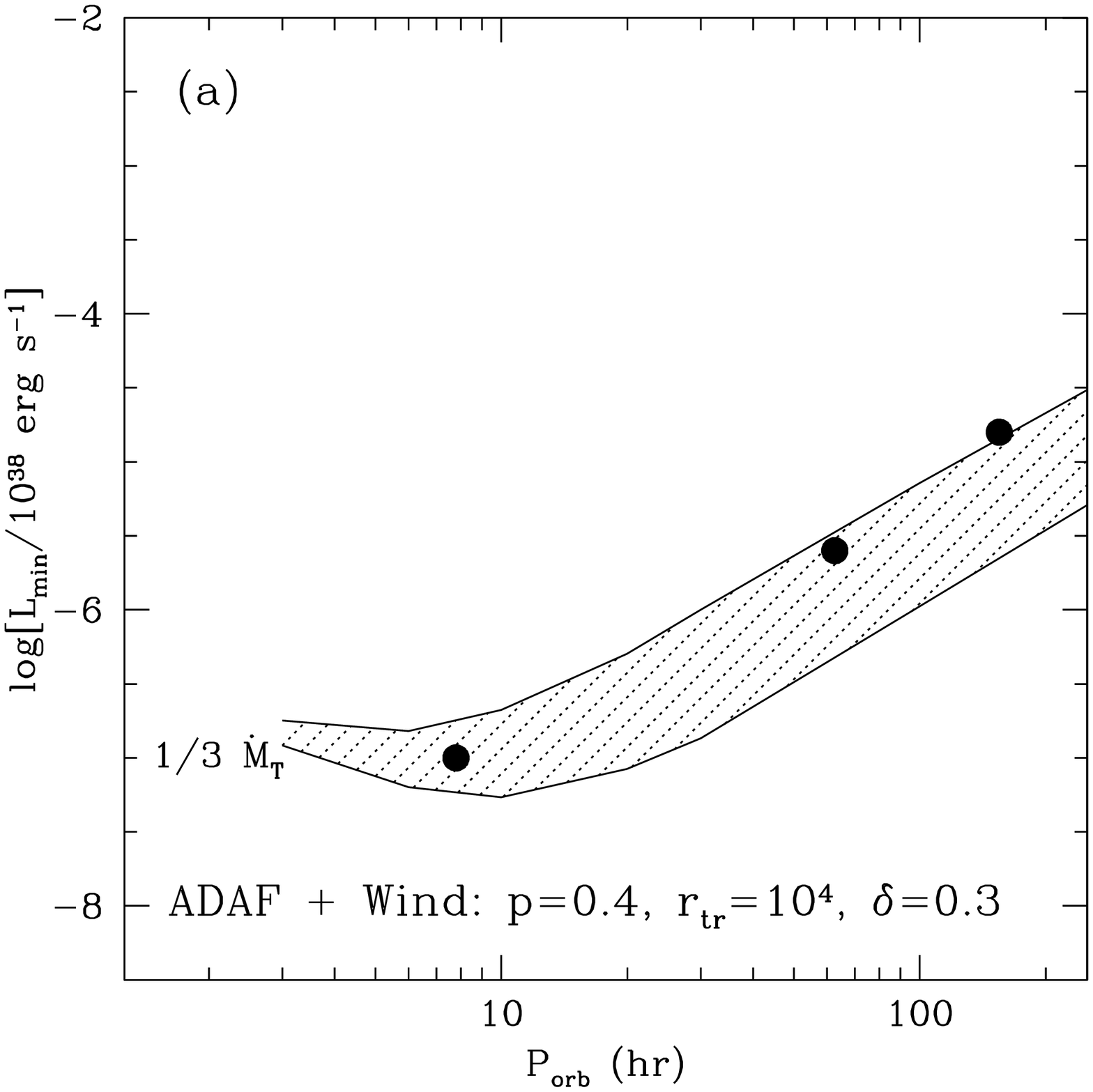}{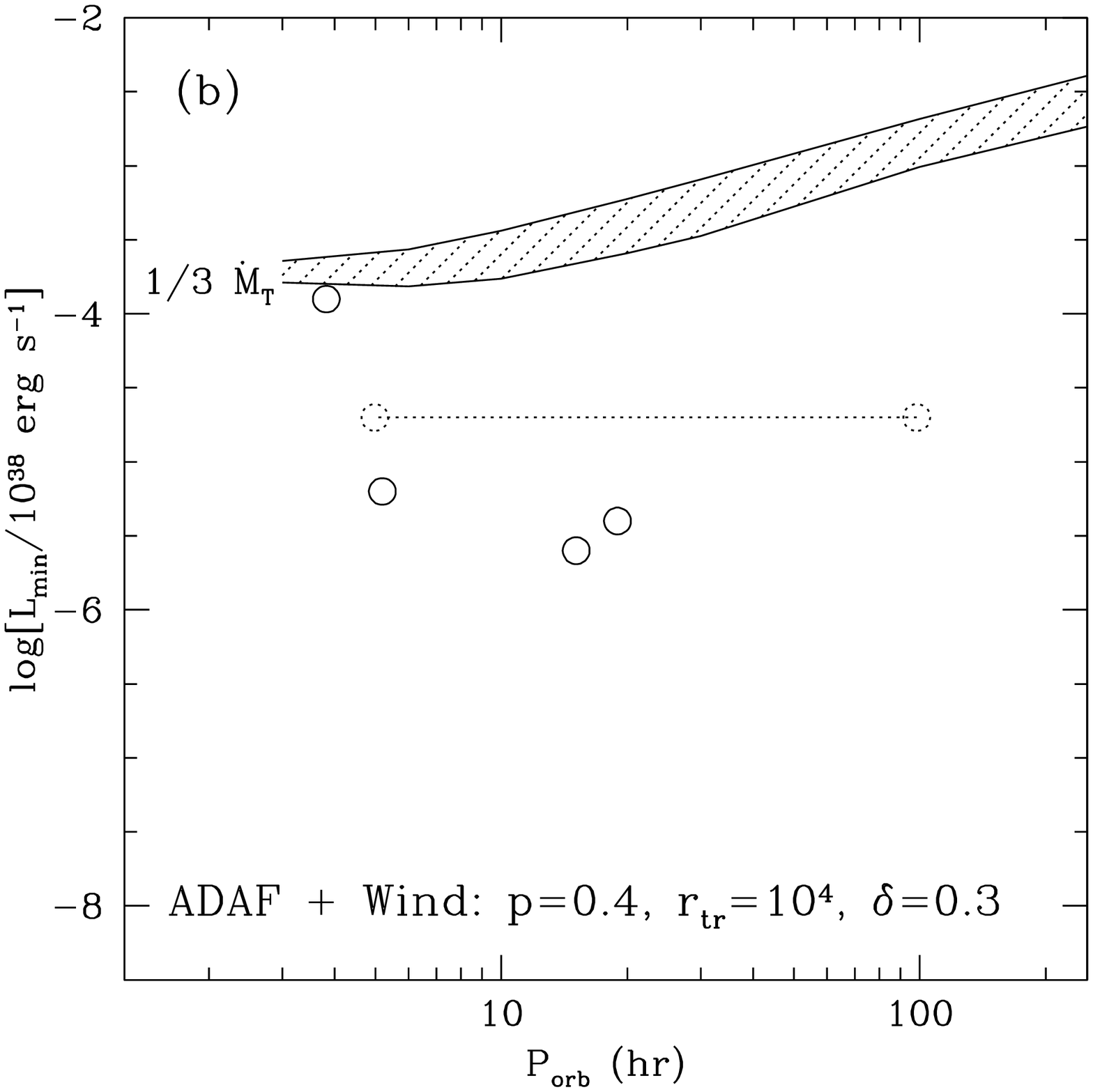}
\caption{(a) The band shows the quiescent luminosities of BH SXTs
  in the 0.5-10 keV band predicted by ADAF models including a wind
  ($p=0.4$, $r_{\rm tr}=10^4$, $\delta=0.3$) if $1/3$ of the mass
  transferred by the secondary flows into the ADAF. (b) The band shows
  the quiescent luminosities of NS SXTs predicted in the 0.5-10 keV
  band if (1) $1/3$ of the mass transferred by the secondary flows
  into the ADAF, (2) there is a modest wind ($p=0.4$, $r_{\rm
    tr}=10^4$, $\delta=0.3$) and (3) all the mass reaching the NS
  magnetosphere is accreted onto the surface. The inner radius of the
  ADAF ($=$ the magnetospheric radius) is at $r_{\rm mq}=20$ (see
  text). The wind reduces, but only by $\sim$ one order of magnitude,
  the accretion rate at the magnetosphere (compare with Fig.~4b). We
  need a modest propeller to decrease the luminosity further to the
  observed levels.
  \label{fig:bhwind}}
\end{figure}

\clearpage

\vspace{2cm}
\begin{table*}
\caption{NEUTRON STAR AND BLACK HOLE SXTs}
\begin{center}
\begin{tabular}{lcccc} \hline \hline
\\

System & $P_{\rm orb}$ (hr) & D (kpc)& $\log [L_{\rm min}]~({\rm
erg~s^{-1}})$& $m_1 (M_{\odot})$ \\ \\ (1) & (2) & (3) & (4) & (5) \\
\\ \hline \\

$\circ$ EXO 0748-676 & $3.82^a$ & $10^e$ & $34.1$& $1.4^{\star}$\\
$\circ$ 4U 2129+47 & $5.2^b$ & $6.3^e$ & $32.8$&$1.4^{\star}$\\
$\circ$ 1456-32 (Cen X-4) & $15.1^a$ & $1.2^e$ &$32.4$&$1.4^{\star}$\\ 
$\circ$ 1908+005 (Aql X-1) & $19^a$ & $2.5^e$ &$32.6$ &$1.4^{\star}$\\ 
$\circ$ H1608-52 & $98.4^c$ or $5^d$ & $3.6^e$& $33.3$&$1.4^{\star}$\\

\\
\hline
\\

$\bullet$ GRO J0422+32 (XN Per 92) & $5.1$ & $2.6^g$& $< 31.6$&$12^k$ \\
$\bullet$ A0620-00 (XN Mon 75) & $7.8$ & $1^h$ & $31.0$& $6.1^h$\\
$\bullet$ GS2000+25 (XN Vul 88)& $8.3$ & $2.7^h$ & $< 32.3$& $8.5^l$\\
$\bullet$ GS1124-683 (XN Mus 91)& $10.4$ & $5^i$& $< 32.4$&$6^i$ \\
$\bullet$ H1705-250 (XN Oph 77)& $12.5$ & $8.6^h$& $< 33.0$& $4.9^l$\\
$\bullet$ 4U 1543-47 & $27.0$ & $8^e$ & $< 33.3^n$& $7^m$\\
$\bullet$ J1655-40 (XN Sco 94)& $62.9$ & $3.2^j$ & $32.4$& $7^j$\\
$\bullet$ GS2023+338 (V404 Cyg)& $ 155.3$ & $3.5^h$ &$33.2$& $12^h$\\

\\
\hline

\end{tabular}
\label{tab:systems}
\end{center}
NOTE. -- (1) $\circ$ indicates a NS primary and $\bullet$ a BH
primary.  (2) Orbital periods from McClintock (1998), except where
indicated.  (3) Distances to the systems.  (4) Luminosities in
quiescence in the 0.5-10 keV band (corrected for the revised
distances) from Narayan et al. (1997b) and Garcia et al. (1998),
except where indicated.  Quiescent luminosities for 4U1543-47 and GRO
J1655-40 are based on 20.4~ks and 100~ks ASCA observations
(respectively), and the quiescent luminosity for GRO J0422+32 is based
on a 19~ks ROSAT observation. (5) Primary masses.  (a) Van Paradijs
(1995).  (b) Simbad CDS Catalog.  (c) Ritter \& Kolb (1998).  (d) Chen
et al. (1998).  (e) Garcia et al. (1998). (g) Esin et al. (1998).  (h)
Narayan et al. (1997b).  (i) Esin et al. (1997).  (j) Hameury et
al. (1997).  (k) Beekman et al. (1997).  (l) Chen et al. (1997).  (m)
This is an arbitrary choice in the range $2.9-7.5$ given by Orosz et
al. (1998; see also Bailyn et al. 1997). (n) Orosz et
al. (1998). ($\star$) For simplicity, all NS masses are assumed to be
$1.4~M_{\odot}$.
\end{table*}


\clearpage


\begin{thebibliography}{}
\bibitem[]{}Abramowicz, M., Chen, X.-M., Granath, M. \& Lasota, J.-P. 1996, ApJ, 471, 762.
\bibitem[]{}Abramowicz, M., Chen, X.-M., Kato, S., Lasota, J.-P., \& Regev, O. 1995, ApJ Lett., 438, L37.
\bibitem[]{}Asai, K. et al. 1998, PASJ, 50, 611.
\bibitem[]{}Avni, Y., Soltan, A., Tananbaum, H. \& Zamorani, G. 1980, ApJ,  238, 800.
\bibitem[]{}Bailyn, C.D., Jain, R.K., Coppi, P. \& Orosz, J.A., 1998., ApJ, 499, 367. 
\bibitem[Beekman et al. (1997)]{beeetal97}Beekman, G., et al., 1997, MNRAS, 290, 303.
\bibitem[]{}Bisnovatyi-Kogan, G.S. \& Lovelace, R.V.E., 1997, ApJ Lett., 486, L43.
\bibitem[]{}Blandford, R.D. \& Begelman, M.C., 1998, MNRAS, submitted,  Astro-ph/9809083 (BB98).
\bibitem[]{}Bondi, H., 1952, MNRAS, 112, 195.
\bibitem[]{}Brown, E.F., Bildsten, L. \& Rutledge, R.E., 1998, ApJ Lett., 504, L95.
\bibitem[]{}Campana, S., Mereghetti, S., Stella, L. \& Colpi, M., 1997, A\&A, 324, 941.
\bibitem[]{}Campana, S. et al., 1998, ApJ Lett., 499, L65.
\bibitem[]{}Cannizzo, J.K. 1993, in Accretion Disks in Compact Stellar Systems, ed. J.C. Wheeler (Singapore: World Scientific), p. 6.
\bibitem[Chen et al. (1997)]{cheshr97}Chen, W., Shrader, C.R. \& Livio, M., 1997, ApJ, 491, 312.
\bibitem[]{}Chakrabarty, D. \& Morgan, E.H., 1998, Nature, 394, 346 (CM98).
\bibitem[]{}Chen, W. et al., 1998, in Accretion Processes in Astrophysics - Some Like it Hot, eds. S. Holt \& T. Kallman (Woodbury, NY: AIP), p. 347.
\bibitem[]{}Chen, X.M., Abramowicz, M.A. \& Lasota, J.-P., 1997, ApJ, 476, 61
\bibitem[]{}Chevalier, C., Ilovaisky, S.A., Van Paradijs, J., Pedersen, H. \& Van der Klis, M., 1989, A\&A, 210, 114.
\bibitem[]{}Cowley, A.P. et al., 1988, AJ, 95, 1231.
\bibitem[]{}Cui, W., 1997, ApJ Lett., 482, L163.
\bibitem[]{}Cui, W., Barret, D., Zhang, S.N., Chen, W., Boirin, L. \& Swank, J., 1998, ApJ Lett., 502, L49.
\bibitem[]{}Daumerie, P.R., 1996, Ph. D. Thesis, Univers. Illinois at Urbana-Champaign. 
\bibitem[]{daf} Davies, R. E., Fabian, A. C. \& Pringle, J. E., 1979, \mnras, 186, 779.
\bibitem[]{dap} Davies, R. E. \& Pringle, J. E., 1981, \mnras, 196, 209. 
\bibitem[Esin et al. (1997)]{esietal97}Esin, A.A., McClintock, J.E., \& Narayan, R., 1997, ApJ, 489, 865.
\bibitem[]{}Esin, A.A., Narayan, R., Cui, W., Grove, J.E. \& Zhang, S.-N., 1998, ApJ, in press, Astro-ph/9711167.
\bibitem[]{}Frank, J., King, A. \& Raine, D., 1992, Accretion Power in Astrophysics (Cambridge: Cambridge University Press).
\bibitem[Garcia et al. (1998)]{garmac97}Garcia, M.R., McClintock, J.E., Narayan, R., \& Callanan, J., 1998, in the Procceedings of the 13th NAW on CVs, Jackson Hole, WY, eds. S. Howell, E. Kuulkers, and C. Woodward, p. 506, Astro-ph/9708149.
\bibitem[]{gho} Ghosh, R. \& Lamb, F. K., 1979, \apj, 234, 296. 
\bibitem[]{}Gruzinov, A., 1998, ApJ, 501, 787.
\bibitem[Hameury et al. (1997)]{hamlas97}Hameury, J.-M., Lasota, J.-P., McClintock, J.E., \& Narayan, R., 1997, ApJ, 489, 234.
\bibitem[]{}Henrichs, H.F. 1983, in Accretion Driven Stellar X-ray Sources, eds. W.H.G. Lewin \& E.P.J. van den Heuvel (Cambridge: Cambridge University Press), p. 393.
\bibitem[]{}Hertz, P., Wood, K.S. \& Cominsky, L.R., 1997, ApJ, 486, 1000. 
\bibitem[]{}Ichimaru, S., 1977, ApJ, 214, 840.
\bibitem[]{ill} Illarionov, A. F. \& Sunyaev, R. A., 1975, A\&A, 39, 185.
\bibitem[]{}Kalogera, V., Kolb, U. \& King, A., 1998, ApJ, in press, Astro-ph/9803288.
\bibitem[]{}King, A., 1988, Quat. Journ. Roy. Astron. Soc., 29, 1.
\bibitem[]{}King, A., Kolb, U. \& Burderi, L., 1996, ApJ Lett., 464, L127.
\bibitem[]{}Kolb, U., 1998, MNRAS, 297, 419.
\bibitem[]{}Kuulkers, E., 1998, New Astron. Rev., in press, Astro-ph/9805031.
\bibitem[]{}Lasota, J.-P. 1996, in Compact Stars in Binaries; IAU Symposium 165, eds. J. van Paradijs  E.P.J. van den Heuvel \& 
         E. Kuulkers, (Dordrecht: Kluwer), p. 43.
\bibitem[]{}Lasota, J.-P. \& Hameury, J.-M., 1998, in Accretion Processes in Astrophysics - Some Like it Hot, eds. S. Holt \& T. Kallman (Woodbury, NY: AIP), p. 351.
\bibitem[]{}Lasota, J.-P., Narayan, R. \& Yi, I., 1996, A\&A, 314, 813.
\bibitem[]{}Lewin, W.H.G., Van Paradijs, J. \& Taam, R.E., 1993, Space Sci. Rev., 62, 223.
\bibitem[]{}McClintock, J.E., 1998, in Accretion Processes in Astrophysics - Some Like it Hot, eds. S. Holt \& T. Kallman (Woodbury, NY: AIP), p. 290. 
\bibitem[]{}McClintock, J.E. \& Remillard, R.A., 1990, ApJ, 350, 386.
\bibitem[]{}Menou, K., Narayan, R. \& Lasota, J.-P., 1998, ApJ, in press.
\bibitem[]{}Meyer, F. \& Meyer-Hofmeister, E., 1994, A\&A, 288, 175.
\bibitem[]{}Nakamura, K.E., Kusunose, M., Matsumoto, R. \& Kato, S., 1997, PASJ, 49, 503.
\bibitem[Narayan et al. (1997)]{narbar97}Narayan, R., Barret, D., \& McClintock, J.E., 1997a, ApJ, 482, 448.
\bibitem[]{nar} Narayan, R., Garcia, M.R. \& McClintock, J.E., 1997b, \apj, 478, L79.
\bibitem[]{}Narayan, R., Kato, S. \& Honma, 1997, ApJ, 476, 49.
\bibitem[]{}Narayan, R., Mahadevan, R., Grindlay, J.E., Popham, R.G. \& Gammie, C.F., 1998a, ApJ, 492, 554.
\bibitem[]{}Narayan, R., Mahadevan, R. \& Quataert, E., 1998b, in The Theory of Black Hole Accretion Discs, eds. M. A. Abramowicz, G. Bjornsson, and J. E.
       Pringle (Cambridge: Cambridge University Press), Astro-ph/9803141.
\bibitem[Narayan et al. (1995)]{mny95}Narayan, R., McClintock, J.E. \&
Yi, I., 1996, ApJ, 457, 821.
\bibitem[]{}Narayan, R. \& Yi, I., 1994, ApJ Lett., 428, L13.
\bibitem[]{}Narayan, R. \& Yi, I., 1995a, ApJ, 444, 231.
\bibitem[]{}Narayan, R. \& Yi, I., 1995b, ApJ, 452, 710.
\bibitem[]{}Orosz, J.A. \& Bailyn, C.D., 1997, ApJ, 477, 876.
\bibitem[]{}Orosz, J.A., Jain, R.K., Bailyn, C.D., McClintock, J.E. \& Remillard, R.A., 1998, ApJ, 499, 375.
\bibitem[]{}Phillips, S.N., Shahbaz, T. \& Podsiadlowski, P. 1999, MNRAS, in press, Astro-ph/9811474.
\bibitem[]{}Popham, R.G. \& Gammie, C.F., 1998, ApJ, in press, Astro-ph/9802321. 
\bibitem[]{}Pylyser, E. \& Savonije, G.J., 1988, A\&A, 191, 57.
\bibitem[]{}Quataert, E., ApJ, 500, 978.
\bibitem[]{}Quataert, E. \& Gruzinov, A., 1998, ApJ submitted, Astro-ph/9803112.
\bibitem[]{}Quataert, E. \& Narayan, R., 1998, ApJ, in press. 
\bibitem[]{}Rajagopal, M. \& Romani, R.W., 1996, ApJ, 461, 327.
\bibitem[]{}Ritter, H. \& Kolb, U., 1998, A\&AS, in press.
\bibitem[]{}Rutledge, R.E., Bildsten, L., Brown, E.F., Pavlov, G.G. \& Zavlin, V.E., 1998, ApJ, in press.
\bibitem[]{}Schmitt, J.H.M., 1985, ApJ, 293, 178.
\bibitem[]{}Shapiro, S.L. \& Salpeter, E.E., 1975, ApJ, 198, 671.
\bibitem[Shaviv \& Wehrse, 1986]{sw86} Shaviv G., Wehrse R., 1986, A\&A Lett., 159, L5.
\bibitem[]{ste}Stella, L., Campana, S., Colpi, M., Mereghetti, S. \& Tavani, M., 1994, ApJ Lett., 423, L47.
\bibitem[]{}Stella, L., White, N.E., Rosner, R., 1986, ApJ, 308, 669.
\bibitem[]{}Tanaka, Y. \& Lewin W.H.G., 1995, in X-ray Binaries, eds. W.H.G. Lewin, J. van Paradijs \& E.P.J. van den Heuvel, (Cambridge: Cambridge University Press), p. 126.
\bibitem[]{}Tanaka, Y. \& Shibazaki, N., 1996, ARA\&A, 34, 607.
\bibitem[]{}Turolla, R., Zampieri, L., Colpi, M. \& Treves, A. 1994, ApJ Lett., 426, L35.
\bibitem[]{}van Paradijs, J., 1995, in X-ray Binaries, eds. W.H.G. Lewin, J. van Paradijs \& E.P.J. van den Heuvel, (Cambridge: Cambridge University Press), p. 536.
\bibitem[]{}van Paradijs, J. \& McClintock, J.E., 1995, in X-ray Binaries, eds. W.H.G. Lewin, J. van Paradijs \& E.P.J. van den Heuvel, (Cambridge: Cambridge University Press), p. 58.
\bibitem[]{ver} Verbunt, F., Belloni, T., Johnston, H.M., van der Klis, M. \& Lewin, W.H.G., 1994, A\&A, 285, 903.
\bibitem[]{}Verbunt, F. \& van den Heuvel, E.P.J., 1995, in X-ray Binaries, eds. W.H.G. Lewin, J. van Paradijs \& E.P.J. van den Heuvel, (Cambridge: Cambridge University Press), p. 457.
\bibitem[]{}Wang, Y.-M., 1987, A\&A, 183, 257.
\bibitem[]{}Wang, Y.-M., 1995, ApJ Lett., 449, L153.
\bibitem[]{wan} Wang, Y.-M. \& Robertson, J. A., 1985, A\&A, 151, 361.
\bibitem[]{}Warner, B., 1995, Cataclysmic Variable Stars, (Cambridge: Cambridge University Press).
\bibitem[]{}Webbink, R.F., Rappaport, S. \& Savonije, G.J., 1983, ApJ, 270, 678.
\bibitem[]{}White, N.E., Nagase, F. \& Parmar, A.N., 1995, in X-ray Binaries, eds. W.H.G. Lewin, J. van Paradijs \& E.P.J. van den Heuvel, (Cambridge: Cambridge University Press), p. 1.
\bibitem[]{}White, N.E. \& Zhang, W., 1997, ApJ Lett., 490, L87.
\bibitem[]{}Wijnands, R. \& van der Klis, M., 1998, Nature, 394, 344. 
\bibitem[]{}Zavlin, V.E., Pavlov, G.G. \& Shibanov, Y.A., 1996, A\&A, 331, 821.
\bibitem[]{}Zhang, S.N., Yu, W. \& Zhang, W., 1998, ApJ Lett., 494, L71.


\end{thebibliography}
\end{document}